\documentclass[reprint,
superscriptaddress,
amsmath,
amssymb,
aps,
prl,
floatfix,
english
]{revtex4-2}

\usepackage{bibunits}
\defaultbibliography{references}
\defaultbibliographystyle{apsrev4-2}
\usepackage{graphicx}% Include figure files
\usepackage{svg}
\usepackage{dcolumn}% Align table columns on decimal point
\usepackage{bm}% bold math
\usepackage{siunitx}
\usepackage[T1]{fontenc}
\usepackage[utf8]{inputenc}

\usepackage[unicode=true, colorlinks=true, citecolor={blue!80!black}, urlcolor={blue!50!black}, linkcolor = {blue!80!black}]{hyperref}

\newcommand{\tdeg}{{\rm 2DEG}}
\newcommand{\SC}{{\rm SC}}
\newcommand{\T}{{\rm T}}
\newcommand{\prox}{{\rm prox}}
\newcommand{\M}{{\rm M}}
\newcommand{\QH}{{\rm QH}}
\newcommand{\lprox}{L}
\newcommand{\kf}{k_\mu}
\newcommand{\pf}{p_F}
\newcommand{\lA}{l_{\rm A}}
\newcommand{\dv}{d}
\newcommand{\Cjump}{{\cal C}_{\rm jump}}

\newcommand{\he}{A_{\rm h}}
\newcommand{\ee}{A_{\rm e}}
\newcommand{\psie}{\eta}
\newcommand{\edg}{{\rm edge}}
\newcommand{\xic}{\xi}

\newcommand{\rngl}{\rangle\hspace{-0.05cm}\rangle}
\newcommand{\lngl}{\langle\hspace{-0.05cm}\langle}

\graphicspath{{Figures/}}

\begin{document}
\widetext

\title{
Disorder in Andreev reflection of a quantum Hall edge
}

\author{Vladislav D.~Kurilovich}
\email{vlad.kurilovich@yale.edu}
\affiliation{Department of Physics, Yale University, New Haven, CT 06520, USA}
\author{Zachary M.~Raines}
\affiliation{Department of Physics, Yale University, New Haven, CT 06520, USA}
\author{Leonid I.~Glazman}
\affiliation{Department of Physics, Yale University, New Haven, CT 06520, USA}

\begin{abstract}

We develop a theory of charge transport along the quantum Hall edge proximitized by a ``dirty'' superconductor. 
Disorder randomizes the Andreev reflection rendering the conductance of a proximitized segment a stochastic quantity with zero average for a sufficiently long segment. We find the statistical distribution of the conductance and its dependence on electron density, magnetic field, and temperature.
\end{abstract}

\maketitle
\textit{Introduction.}---
Recent interest in engineering an exotic superconductor have renewed the effort to combine the superconducting proximity effect with a quantizing magnetic field. The combination of the two
has been proposed as a route to realize new quasiparticles, such as parafermions~\cite{mong2014, clarke2014}, which may be employed for topological quantum computing~\cite{nayak2008}. 
The picture of the proximity effect is based on Andreev reflection, in which an electron incident on the interface between a normal state conductor and a superconductor is
reflected as a hole~\cite{andreev1964}. 
In fact, this electron-hole conversion has been demonstrated~\cite{bozhko1982,benistant1983} in focusing experiments utilizing a weak magnetic field $B$ to bend the electron and hole trajectories.  
Classically, trajectory bending due to the Lorentz force leads to formation of skipping orbits propagating along the boundaries. 
At fixed energy, quasiclassical quantization results in a discrete spectrum of 
angles $\alpha_n(B)$ such a trajectory may form with the boundary, varying continuously with $B$.  
For electron-hole conversion at a boundary with a {\it clean} superconductor, the angles of incidence and reflection obey the
{\it retroreflection} condition, $\alpha_n(B)+\alpha_m(B)=\pi$.
Clearly, the latter is satisfied only for a discrete set of fields $B$.
In weak fields, {\it i.e.,} at high filling factors $\nu\gg 1$, this set is dense,
and one may disregard the consequence of its discreteness~\cite{eroms05, batov2007}.

The described electron and hole ``magnetic surface levels''~\cite{nee1967} are known as the edge states in the context of the quantum Hall effect. The angle matching problem becomes severe for a smaller $\nu$.
For
a single edge state ($\nu = 2$), the matching condition is satisfied only for one specific value of $B$. Electron-hole conversion is effective only at that fine-tuned value of the magnetic field.

Disorder, however, lifts the retroreflection constraint, and allows for an appreciable electron-hole conversion at all magnetic fields.
Indeed, a strong conversion signal was observed in recent experiments~\cite{lee2017,zhao2020,gul2021,hatefipour2021} without any fine tuning; the need of high critical fields $H_{\rm c 2}$ dictated the use of ``dirty'' superconductors.
Robust Andreev reflection, being enabled by disorder, is naturally sensitive to its realization in a sample.
As a result, the charge transport varies stochastically with control parameters such as the magnetic field or the electron density, 
as is observed both in experiment~\cite{zhao2020} and in numerical simulation~\cite{manesco21}.

The crucial difference of conduction along the proximitized quantum Hall edge from the conventional mesoscopic transport stems from the chirality of the edge states.
This renders the well-established theory of mesoscopic conductance fluctuations~\cite{altshuler1985, lee1985} inapplicable.
In this work, we develop a quantitative theory of mesoscopic quantum transport along the proximitized chiral edge, making predictions for the statistics of conductance fluctuations and their dependence on electron density, magnetic field, and temperature. 
The results obtained for chiral transport differ substantially from their counterpart in usual conductors. 

\begin{figure}[t]
  \begin{center}
    \includegraphics[scale = 1]{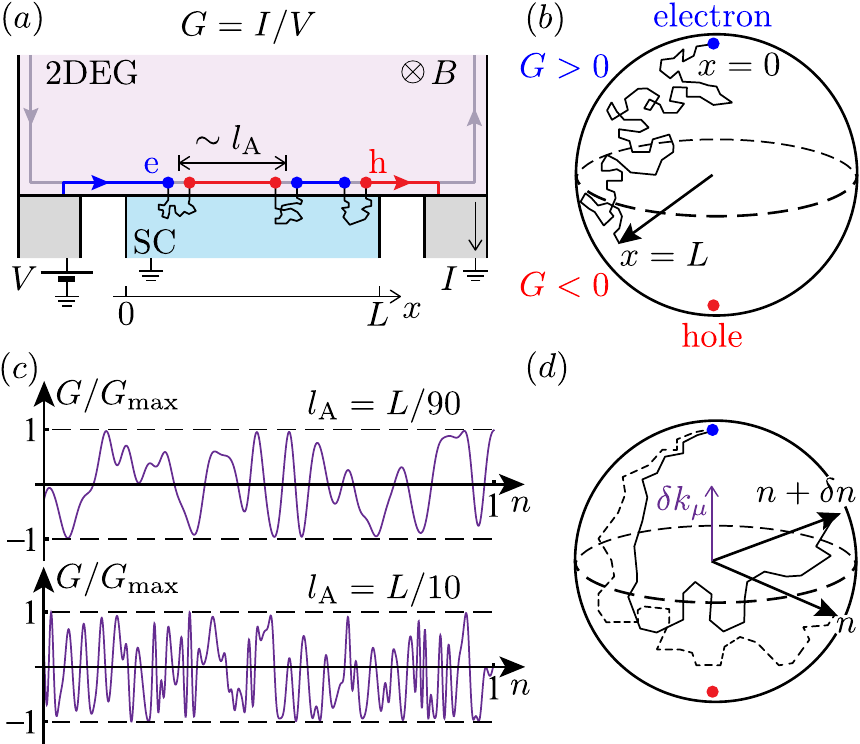}
    \caption{
    (a) A chiral edge state with a segment proximitized by a ``dirty'', grounded superconductor. Electrons are launched towards the segment from an upstream electrode biased by voltage $V$. An electron propagating along the segment converts randomly into a hole over the distance $\lA$, which is controlled by disorder in the superconductor, see Eq.~\eqref{eq:var}.
    (b) Evolution of the electronic wave function, see Eq.~\eqref{eq:evol}, is similar to the motion of a ``spin'' in a stochastic effective ``magnetic field".
    The conductance $G = I / V$ is determined by the result of a random walk of a point on a Bloch sphere.
    (c) $G$ is a random quantity that fluctuates upon varying the electron density $n$ in the 2DEG (traces are simulated using Eq.~\eqref{eq:evol}; units of $n$ are the same for the two plots and are otherwise arbitrary).
    (d) The loss of correlation between the values of $G$ upon a change in $n$ is quantified by function ${\cal C}(\delta n)$, see Eqs.~\eqref{eq:corrdef}--\eqref{eq:ncor}. The origin of the correlations loss is illustrated by the divergence between two stochastic trajectories on a Bloch sphere.
    The ``spins'' corresponding to different values of $n$ experience a different effective ``magnetic field'', and thus drift apart in the course of evolution.
    The separation of the ``spins'' is slower for stronger disorder.
    As the result, the trace $G(n)$ in panel (c) is smoother for smaller $\lA$.
    \label{figure}}
  \end{center}
\end{figure}

\textit{Model.}---We are interested in the linear conductance $G$ in a three-terminal setting, see Fig.~\ref{figure}(a). 
To find $G$, we start with the Hamiltonian
\begin{equation}
    H = H_\tdeg + H_\SC + H_\T.
\end{equation}
Here, $H_\tdeg$ describes the two-dimensional electron gas (2DEG) in a $\nu = 2$ quantum Hall state. $H_\SC$ is the Hamiltonian of the superconductor.
We consider the experimentally relevant~\cite{lee2017,zhao2020,gul2021,hatefipour2021} ``dirty'' limit 
$l_{\rm mfp} \ll \xic$, where $l_{\rm mfp}$ and $\xic$ are, respectively, the electron mean free path and the coherence length in the superconductor.
Coupling between the 2DEG and superconductor is described by the tunneling Hamiltonian~\cite{prada2004, lutchyn2012}
\begin{equation}\label{eq:HT}
    H_\T = t \sum_{\sigma} \int_0^L dx\, (\partial_y \psi^\dagger_\sigma(x,0)\partial_y \chi_\sigma (x,0,0) + {\rm h.c.}),
\end{equation}
where $\psi_\sigma (x, y)$ and $\chi_\sigma (x,y,z)$ are annihilation operators for an electron with spin $\sigma =\,\uparrow$ or $\downarrow$ in the 2DEG and superconductor, respectively. 
The interface of length $L$ is located at $y = z = 0$. 
For simplicity, we assume that the tunneling amplitude $t$ is uniform along the interface. 

For the purpose of describing transport at low temperature and bias, it is convenient to derive an effective Hamiltonian focusing on chiral electrons at the 2DEG's edge,
\begin{equation}\label{eq:eff}
    H_{\rm eff} = H_\edg + H_\prox.
\end{equation}
The first term is obtained by projecting $H_\tdeg$ onto the subspace of edge states belonging to a single Landau level
\begin{equation}
    H_\edg = \sum_\sigma \int dx\,\psie^\dagger_\sigma(x) \hbar v[-i\partial_x - \kf]\psie_\sigma(x).
\end{equation}
Here, $\eta_\sigma(x)$ is a field operator for chiral electrons with $\sigma = \,\uparrow$ or $\downarrow$, $v$ is their velocity, and $k_\mu$ is the Fermi momentum; we neglect the Zeeman splitting. The second term in Eq.~\eqref{eq:eff} describes the effect of superconducting proximity. It is obtained by a standard Schrieffer-Wolff transformation.
For electron energies $E \ll \Delta$, as measured from the Fermi level, the transformation results in
\begin{equation}
    H_\prox = (\partial_y \Phi)^2 t^2\hspace{-0.10cm}\int_0^L\hspace{-0.10cm} dx_1 dx_2\, \hat{\psie}^\dagger(x_1) \partial^2_{y_1 y_2} {\cal G} (x_1,x_2) \hat{\psie}(x_2),
\end{equation}
where $\hat{\psie}(x) = (\psie_\uparrow (x),\,-\psie_\downarrow^\dagger(x))^T$, the $2\times 2$ matrix ${\cal G}(x_1, x_2)$ is the Green's function of the superconductor at $E = 0$ (arguments $y_{1,2}, z_{1,2} = 0$ are suppressed for brevity), $\Delta$ is the energy gap in the superconductor, and $\Phi(y)$ is the transverse component of the edge state's wave function at the Fermi level.

Conductance $G$ at $T = 0$ can be expressed in terms of transmission amplitudes across the proximitized segment in the normal ($\ee$) and Andreev ($\he$) channels
at $E = 0$,
\begin{equation}\label{eq:G}
    G = G_Q (|\ee|^2 - |\he|^2),
\end{equation}
where $G_Q = 2e^2 / h$ is the conductance quantum.
To find $G$ in the setup of Fig.~\ref{figure}(a), we thus need to solve a quantum-mechanical scattering problem.

\textit{Andreev amplitude for a short segment.---}An electron experiences at most one Andreev reflection  while propagating along a sufficiently short proximitized segment. The corresponding Andreev amplitude can be found perturbatively in $H_{\rm prox}$. With the help of Born approximation, we obtain
\begin{equation}\label{eq:damp}
    \he \hspace{-0.1cm} =\hspace{-0.05cm} -\frac{(\partial_y \Phi)^2 t^2}{v}\hspace{-0.2cm}\int \hspace{-0.1cm} dx_1 dx_2 e^{i\kf (x_1 + x_2)}\partial^2_{y_1 y_2}{\cal G}_{\rm he} (x_1, x_2),
\end{equation}
where ${\cal G}_{\rm he}$ is the anomalous component of the superconductor Green's function.

The Green's function in Eq.~\eqref{eq:damp} is determined by the interference of electron waves in the superconductor.
The stochastic interference pattern is sensitive to a particular disorder landscape in the region of size $\sim \xic$ adjacent to the interface. Thus, ${\cal G}_{\rm he}$ and $\he$ of Eq.~\eqref{eq:damp} are random quantities.
The latter fluctuates upon varying the magnetic field or the electron density in the 2DEG.

To characterize the statistical properties of the amplitude, we first find $\langle \he \rangle$.
The averaging here is performed over a sufficiently broad window of magnetic fields or electron densities.
Formally, it is equivalent to averaging over the possible disorder configurations in the superconductor.
With the help of the latter, more practical definition we obtain:
$\langle \he \rangle \propto \int dx_1 dx_2 e^{ik_\mu (x_1 +x_2)} \langle {\cal G}_{\rm he} (x_1 - x_2)\rangle \propto \int dx e^{2ik_\mu x} \propto \delta(k_\mu)$.
We see that $\langle \he \rangle = 0$ unless $k_\mu = 0$. In the following, we disregard such a fine-tuning and take $\langle \he \rangle = 0$.

Next, we compute the average probability of the Andreev reflection $\langle |\he|^2\rangle$.
As follows from Eq.~\eqref{eq:damp}, we need to average product of the anomalous Green's functions of the superconductor. 
Such an average can be expressed in terms of the normal-state diffuson and Cooperon via a standard procedure 
(see, e.g., Ref.~\onlinecite{hekking1994}).
Assuming that the thickness of the superconducting film and $\lprox$ exceed $\xic$, we obtain~\cite{sm}
\begin{equation}\label{eq:var}
    \langle |\he|^2 \rangle = \frac{\lprox}{\lA},\quad\quad\quad \frac{1}{\lA} = \frac{4\pi g^2}{G_Q \sigma} \ln \frac{\xic}{l_{\rm mfp}}.
\end{equation}
Here $g = 2\pi^2 G_Q t^2 (\partial_y \Phi)^2   \nu_\QH  \nu_\M \pf / \hbar$ is the conductance per unit length of the interface between the quantum Hall edge and the metal in the normal state. Along with the dependence on $\Phi(y)$, the conductance $g$ is proportional to the one-dimensional density of edge states $\nu_{\rm QH} = 1/(2\pi \hbar v)$.
It is also proportional to the normal-state density of states $\nu_{\rm M}$ and Fermi momentum $p_F$ in the superconductor.
Unlike in the clean case, the leading contribution to the Andreev reflection comes from electron trajectories much longer than the Fermi wavelength, with length scale set instead by $\xic \gg l_{\rm mfp}$.
The presence of the logarithmic factor and the appearance of the normal-state conductivity $\sigma$ in $1/\lA$ results from the diffusive motion of electron in the superconductor.

The perturbative result, Eq.~\eqref{eq:damp}, is applicable at $\lprox \ll \lA$. Under this condition, $A_{\rm h}$ is a Gaussian random variable which allows one to compute all moments of $A_{\rm h}$ distribution. 
Using Eq.~\eqref{eq:G} we find $\langle G \rangle = G_Q ( 1 - 2 L / l_{\rm A})$ and $\lngl G^2 \rngl = \langle G^2\rangle - \langle G\rangle^2 =4G^2_Q L^2 / \lA^2$ for the average value and fluctuation of the conductance.

\textit{Conductance of a long segment.}---At $\lprox \gg \lA$,
an incident electron experiences multiple Andreev reflections upon traversing the proximitized segment. The first-order perturbation theory cannot be applied directly to find the amplitude $A_{\rm h}$ in this case.
Instead, we track how the quasiparticle wave function evolves along the segment piece by piece.

We break the segment into a series of short elements with length $\delta L$ satisfying $\xic \ll \delta L \ll \lA$. Under these conditions, the Andreev amplitudes of different elements $\delta \he (x)$ are statistically independent and may still be evaluated perturbatively,
$\delta \he(x) = \alpha(x) \cdot \sqrt{\delta L}$. In addition to Andreev reflections, a quasiparticle may experience forward scattering due to an excursion in the superconductor. 
Similarly to $\delta \he (x)$, we find~\cite{sm} for the electron forward scattering phase $\delta \Theta (x) = \vartheta(x)\cdot \sqrt{\delta L}$. Variables $\alpha(x)$ and $\vartheta(x)$ are Gaussian and independent, $\langle \alpha(x) \vartheta (x^\prime)\rangle = 0$. Using Eq.~\eqref{eq:var} and a similar relation for $\langle \Theta^2 \rangle$ we obtain for the correlators
\begin{equation}\label{eq:corrs}
    \langle \alpha(x) \alpha^\star (x^\prime) \rangle = \langle \vartheta (x) \vartheta (x^\prime) \rangle = \frac{1}{\lA} \delta (x - x^\prime).
\end{equation}
The change of the wave function across each element is small. Therefore, we can describe the wave function evolution by a differential equation:
\begin{equation}\label{eq:evol}
    i \frac{\partial}{\partial x} \begin{pmatrix}
    a_{\rm e}(x)\\a_{\rm h}(x)
    \end{pmatrix} = 
    \begin{pmatrix}
    -\vartheta(x) & \alpha^\star(x)\\
    \alpha(x) & \vartheta(x)
    \end{pmatrix}
    \begin{pmatrix}
    a_{\rm e}(x)\\a_{\rm h}(x)
    \end{pmatrix}.
\end{equation}
Here $a_{\rm e}(x)$ and $a_{\rm h}(x)$ are the electron and hole components of the quasiparticle wave function, respectively.

Equation~\eqref{eq:evol} describes a unitary evolution of a two-component spinor, which can be visualized as a random walk of a point on a Bloch sphere, see Fig.~\ref{figure}(b).
We parameterize $a_{\rm e}(x) = \cos (\theta(x) / 2)$ and $a_{\rm h}(x) = e^{i\phi(x)} \sin (\theta(x) / 2)$, where $\theta$ and $\phi$ are polar and azimuthal angles on the sphere, respectively~\footnote{In the parameterization of  $a_{\rm e}$ and $a_{\rm h}$, we suppressed the common phase which is inconsequential for $G$.}.
The conductance $G = G_Q \cos \theta(L)$ can be expressed in terms of a solution of Eq.~\eqref{eq:evol} with initial condition $\theta(0) = 0$.

To determine the statistics of conductance fluctuations, we derive a Fokker-Planck equation~\cite{vankampen2007} for the distribution function ${\cal P}(\theta, \phi|x)$ with help of Eq.~\eqref{eq:corrs}:
\begin{equation}\label{eq:FP}
    \frac{\partial {\cal P}}{\partial x} = \frac{1}{\lA}\left(\Delta_{\theta, \phi}  + \partial^2_\phi\right) {\cal P}.
\end{equation}
Here $1/\lA$ plays the role of a diffusion coefficient in the amplitude's random walk.
Equation \eqref{eq:FP} can solved straightforwardly in terms of angular harmonics,
${\cal P}(\theta,\phi|x) = \sum_{l=0}^{\infty}(2l+1) P_l(\cos \theta) e^{-l(l+1)x / \lA} / 4\pi$, where $P_l(z)$ are Legendre polynomials.

Using the found distribution function, we obtain for the average conductance \footnote{Exponential decay of $\langle G \rangle$ with $L$ was also noted in Ref.~\cite{zhang2020}.}:
\begin{equation}\label{eq:mean}
    \langle G \rangle =  G_Q e^{-2\lprox / \lA}.
\end{equation}
At $\lprox \gg \lA$, conductance $G$ is distributed uniformly in the interval $[-G_Q, G_Q]$ with $\langle G \rangle = 0$ and variance $\lngl G^2 \rngl = G^2_Q / 3$. Thus, the conductance fluctuations pattern is sign-alternating and evenly distributed between positive and negative values, see Fig.~\ref{figure}(c).

\textit{Suppression of fluctuations by vortices.}---Only a type II superconductor can withstand magnetic field $B$ required to enter the quantum Hall regime in the 2DEG.
Such field
induces vortices, which lead to a non-vanishing density of states in the superconductor at the Fermi level~\cite{wattstobin1973}.
As a result, an electron or a hole propagating along the edge can tunnel normally into the superconducting electrode thus not contributing to $G$.
This leads to attenuation of conductance fluctuations.

The probability of an incident electron to survive the propagation along the proximitized segment and reach the downstream electrode (as a particle or as a hole) decreases exponentially with $\lprox$:
\begin{equation}\label{eq:surv}
    p_{\rm surv} = \exp \left[-\frac{g\lprox}{G_Q} \frac{\bar{\nu}}{\nu_{\rm M}}\right].
\end{equation}
Here the induced by vortices density of states $\bar{\nu}$ is taken at $E = 0$ and averaged along the interface. 
Despite the attenuation, at $\lprox \gg \lA$ the conductance distribution remains uniform. However, its spread reduces to the interval $[-G_{\rm max}, G_{\rm max}]$ and its variance becomes
\begin{equation}\label{eq:variance}
    \lngl G^2 \rngl = \frac{G_{\rm max}^2}{3},\quad\quad G_{\rm max} = G_Q\, p_{\rm surv}.
\end{equation}
Ratio $\bar{\nu} / \nu_{\rm M}$ in Eq.~\eqref{eq:surv} increases with $B / H_{\rm c 2}$, reaching unity at the upper critical field, $B = H_{\rm c2}$.
Consequently, 
$\lngl G^2 \rngl$ decreases with increasing $B$. This is qualitatively consistent with the observations of Ref.~\cite{zhao2020}.

\textit{Conductance correlation function.}---We now find the correlation function of the conductance fluctuations with the electron density $n$ in the 2DEG,
\begin{equation}\label{eq:corrdef}
    {\cal C}(\delta n) = \lngl G(n) \cdot G(n + \delta n)\rngl.
\end{equation}
Variation of density $\delta n$ shifts the Fermi momentum of chiral electrons by $\delta k_\mu = \delta n (\partial \mu / \partial n) / (\hbar v)$, where $\partial \mu / \partial n$ is the inverse compressibility of the quantum Hall state.
$\delta k_\mu$ affects the phases of Andreev reflection amplitudes, whose interference determines the conductance.
We see from Eq.~\eqref{eq:damp} that $\alpha(x) \rightarrow \alpha(x) e^{2i\delta k_\mu x}$ upon changing $n \rightarrow n + \delta n$.
Applying this modification to Eq.~\eqref{eq:evol} and using Eq.~\eqref{eq:corrs}, we derived~\cite{sm} a differential equation for ${\cal C}(\delta n)$ as a function of $L$. Solving it, we find at $\lprox \gg \lA$: 
\begin{equation}\label{eq:corrf}
    {\cal C}(\delta n) = \lngl G^2 \rngl\, \exp \Big[-\frac{4}{3}\Big(\frac{\delta n}{n_{\rm cor}}\Big)^2\Big].
\end{equation}
The correlation density $n_{\rm cor}$ is given by:
\begin{equation}\label{eq:ncor}
    n_{\rm cor} = \frac{\partial n}{\partial \mu} \frac{\hbar v}{\sqrt{\lA \lprox}}.
\end{equation}
The dependence of Eq.~\eqref{eq:ncor} on $\lprox$ and $\lA$ is of particular note.
Firstly, $n_{\rm cor} \propto 1/\sqrt{L}$ reflects the diffusive character of the wave function evolution.
In contrast, periodic oscillations of the quasiparticle between electron and hole states in the absence of disorder would lead to ${\cal C}(\delta n)$ variation on a scale $\delta n \propto 1/L$~\cite{lian2016}. Secondly, $n_{\rm cor} \propto 1 / \sqrt{\lA}$ increases with disorder in superconductor, as $l_{\rm A} \propto \sigma$, cf.~Eq.~\eqref{eq:var}. 
Thus, the pattern of mesoscopic fluctuations is \textit{smoother} for a dirtier superconductor, see Fig.~\ref{figure}(c). 
This unusual behavior is similar in its origin to the motional narrowing in nuclear magnetic resonance~\cite{slichter1990}.

The conductance also fluctuates with the magnetic field. 
The generalization of Eq.~\eqref{eq:corrf} reads ${\cal C}(\delta n, \delta B) = \lngl G^2 \rngl \exp[-\frac{4}{3} \delta k_{\mu}^2 \lA \lprox] \exp[- \frac{8}{3} (\delta g / g)^2 \lprox / \lA ]$.
Change in the Fermi momentum $\delta k_\mu (\delta n, \delta B)$ varies the phases of the Andreev reflection amplitudes (as discussed above). Variation $\delta g (\delta n, \delta B)$
affects the amplitudes magnitude through the dependence of $\Phi(y)$ and $v$ on $B$ and $n$, cf.~Eq.~\eqref{eq:damp}. The functions $\delta g$ and $\delta k_\mu$ acquire a particularly simple form in the limit of high compressibility \footnote{The high compressibility means 
small disorder-induced broadening of Landau levels, $\delta \varepsilon \ll \hbar \omega_{\rm c}$. 
We also assumed the London penetration depth $\lambda \gg \lA$ to neglect the diamagnetic current effect on $\delta k_\mu$.}:
$\delta g / g = 2\, \delta B / B$ and $\delta k_\mu (\delta n, \delta B) = \frac{1}{v}\frac{\partial \mu}{\partial n} [\delta n - \nu \delta B / \phi_0]$, where $\nu(n, B)$ is the quantum Hall filling factor and $\phi_0 = h c / e$.

\textit{Effect of a vortex entrance.}---
In the above we disregarded the entrance of vortices in the superconductor through the interface.
An entering vortex introduces a kink in the phase of the order parameter near the interface. This affects the interference between the Andreev reflection processes thus leading to a jump $\delta G$ in the conductance.

The magnitude of $\delta G$ is a random quantity whose statistical properties depend on the relation between $\dv$ and $\lA$, where $d$ is the distance of the vortex core to the interface. We compute the variance, $\Cjump(d) = \langle (\delta G)^2 \rangle$,
where the average is evaluated over a window of electron densities of width exceeding $n_{\rm cor}$. To do that, we compare the results of the wave function evolution along the proximitized segment before and after the vortex has entered.

The vortex entrance leads to $\alpha(x) \rightarrow \alpha(x)e^{-i\delta \varphi(x)}$ in Eq.~\eqref{eq:evol}.  
The phase $\delta \varphi(x) = \pi + \arctan ([x - x_{\rm v}]/\dv)$ interpolates between $0$ and $2\pi$ over the interval $|x - x_{\rm v}| \sim d$, where $x_{\rm v}$ is the $x$-coordinate of the vortex core. The overall interference pattern does not change substantially if $\dv \ll \lA$. Under this condition, the conductance jump is small. 
It is also small in the opposite limit, $\dv \gg \lA$, in which the presence of $\delta \varphi(x)$ 
can be accounted for with the help of the adiabatic approximation applied to Eq.~\eqref{eq:evol}. We find~\cite{sm}
\begin{equation}\label{eq:jump}
    \frac{\Cjump(\dv)}{\lngl G^2 \rngl} =
    \begin{cases}
    \frac{32\pi d}{3\lA},&\quad d\ll \lA,\\
    \frac{4\pi \lA}{3d},&\quad d \gg \lA. 
    \end{cases}
\end{equation}
The two asymptotes match each other at $\dv \sim \lA$. In this case, the conductance jump is maximal and comparable to the signal itself, $\Cjump(\dv) \sim \lngl G^2 \rngl$. This regime is relevant for the data presented in Ref.~\cite{zhao2020}.

\textit{Conductance fluctuations at finite temperature.---}
In a conventional mesoscopic conductor, the electron transmission amplitudes at energies $E_1$ and $E_2$ are uncorrelated if $|E_1 - E_2| \gtrsim E_{\rm Th}$.
The Thouless energy here is determined by the electron propagation time across the sample; $E_{\rm Th} = \hbar v / L$ in the ballistic limit.
Thus, the ordinary mesoscopic conductance fluctuations~\cite{altshuler1985,lee1985} are smeared out at temperature $T \gtrsim T_{\rm sm} = \hbar v/ L$.

While quasiparticles propagate ballistically along the proximitized quantum Hall edge,
the energy scale $ \hbar v / L$ is \textit{irrelevant} for the correlation of Andreev amplitudes. 
The main mechanism responsible for the variation of $\he$ with $E$ is the dependence of the anomalous Green function on $E / \Delta$ in Eq.~\eqref{eq:damp} generalized to finite energy \footnote{We assume $\Delta \ll \hbar \omega_{\rm c}$ and disregard other mechanisms controlled by $E / (\hbar \omega_{\rm c})$.}.
Using Eq.~\eqref{eq:evol} to compare $\he$ at different $E$, it is easy to show that fluctuations of $G$ at $\lprox \gg \lA$ are smeared out above $T_{\rm sm} \sim \Delta\,(\lA / \lprox)^{1/4}$, in stark contrast with a conventional ballistic conductor.
The difference stems from
the chiral nature of the edge, which prohibits backscattering and formation of standing waves.

The found weak dependence, $T_{\rm sm} \propto L^{-1/4}$, prompts us to explore inelastic scattering as a mechanism of the fluctuations suppression.
In one dimension, inelastic pair collisions are forbidden by the energy and momentum conservation~\cite{imambekov2012}.
Violation of translation invariance by disorder allows for the pair collisions at the edge and leads to a standard Fermi liquid estimate for the scattering rate, $\tau_{\rm in}^{-1}(T) = b\,T^2$~\cite{kane1995}.
The conductance fluctuations are suppressed at temperature exceeding $T_{\rm in}$ such that $v\,\tau_{\rm in}(T_{\rm in}) \sim L$.
We then find $T_{\rm in} \propto L^{-1/2}$.

The coefficient $b$ is not universal and depends on disorder. If the latter is due to the electron excursions into the dirty superconductor, then we can estimate $b \sim \big[\frac{e^2}{\kappa \hbar v}\big]^2 \frac{v}{\lA} \frac{1}{(\hbar \omega_{\rm c})^{2}}$,
where $\kappa$ is the dielectric constant of the environment and $\omega_{\rm c}$ is the cyclotron frequency. In this case, we obtain $T_{\rm in} \sim \hbar \omega_{\rm c} [\kappa \hbar v / e^2] (\lA/\lprox)^{1/2}$. The comparison of $T_{\rm in}$ and $T_{\rm sm}$ is very sensitive to $\hbar \omega_{\rm c} / \Delta$.

\textit{Conclusions.---}
Disorder allows for efficient Andreev reflection of a quantum Hall edge without fine-tuning, but it
introduces randomness in the edge transport.
Electrons stochastically convert into holes over a
length scale $l_{\rm A}$, see Eq.~\eqref{eq:var}.
This stochasticity 
results in conductance fluctuations with the variation of electron density or magnetic field strength. For a long edge, $\lprox \gg \lA$, the
average conductance $\langle G \rangle$ vanishes, see Eq.~\eqref{eq:mean}, while in the absence of vortices the individual realizations of $G$ vary within an interval $\pm 2e^2/h$. 
Electron tunneling into the cores of the vortices in the superconductor shrinks this interval, see Eqs.~\eqref{eq:surv} and \eqref{eq:variance}, due to electrons being lost to ground.
The ensemble averaging of $G$ can be experimentally achieved in a given sample by varying the electron density $n$ by amount exceeding $n_{\rm cor}$ of Eq.~\eqref{eq:ncor}. At smaller variation, the values of $G$ are correlated, see Eq.~\eqref{eq:corrf}. Variation of magnetic field also results in conductance fluctuations, including abrupt changes associated with a vortex entering the superconductor, see Eq.~\eqref{eq:jump}. At a finite temperature, thermal smearing and inelastic scattering suppress conductance fluctuations.
The chiral nature of edge states, however, weakens the suppression compared to the case of conventional conductors.
Our work explains the basic findings of experiment~\cite{zhao2020} including the observation of random conductance, with zero average. Our quantitative predictions call for further experiments exploring the conductance fluctuations pattern. 

\textit{Acknowledgements.---}We acknowledge very useful discussions with
Ethan G.~Arnault, Meng Cheng, Gleb Finkelstein, Pavel D.~Kurilovich, Felix von Oppen, and
Lingfei Zhao.
This work was supported by NSF DMR-2002275 (V.D.K.~and L.I.G.) and by the Yale Prize Postdoctoral Fellowship in Condensed Matter Theory (Z.M.R.).

\bibliography{references}

\begin{thebibliography}{31}%
\makeatletter
\providecommand \@ifxundefined [1]{%
 \@ifx{#1\undefined}
}%
\providecommand \@ifnum [1]{%
 \ifnum #1\expandafter \@firstoftwo
 \else \expandafter \@secondoftwo
 \fi
}%
\providecommand \@ifx [1]{%
 \ifx #1\expandafter \@firstoftwo
 \else \expandafter \@secondoftwo
 \fi
}%
\providecommand \natexlab [1]{#1}%
\providecommand \enquote  [1]{``#1''}%
\providecommand \bibnamefont  [1]{#1}%
\providecommand \bibfnamefont [1]{#1}%
\providecommand \citenamefont [1]{#1}%
\providecommand \href@noop [0]{\@secondoftwo}%
\providecommand \href [0]{\begingroup \@sanitize@url \@href}%
\providecommand \@href[1]{\@@startlink{#1}\@@href}%
\providecommand \@@href[1]{\endgroup#1\@@endlink}%
\providecommand \@sanitize@url [0]{\catcode `\\12\catcode `\$12\catcode
  `\&12\catcode `\#12\catcode `\^12\catcode `\_12\catcode `\%12\relax}%
\providecommand \@@startlink[1]{}%
\providecommand \@@endlink[0]{}%
\providecommand \url  [0]{\begingroup\@sanitize@url \@url }%
\providecommand \@url [1]{\endgroup\@href {#1}{\urlprefix }}%
\providecommand \urlprefix  [0]{URL }%
\providecommand \Eprint [0]{\href }%
\providecommand \doibase [0]{https://doi.org/}%
\providecommand \selectlanguage [0]{\@gobble}%
\providecommand \bibinfo  [0]{\@secondoftwo}%
\providecommand \bibfield  [0]{\@secondoftwo}%
\providecommand \translation [1]{[#1]}%
\providecommand \BibitemOpen [0]{}%
\providecommand \bibitemStop [0]{}%
\providecommand \bibitemNoStop [0]{.\EOS\space}%
\providecommand \EOS [0]{\spacefactor3000\relax}%
\providecommand \BibitemShut  [1]{\csname bibitem#1\endcsname}%
\let\auto@bib@innerbib\@empty
%</preamble>
\bibitem [{\citenamefont {Mong}\ \emph {et~al.}(2014)\citenamefont {Mong},
  \citenamefont {Clarke}, \citenamefont {Alicea}, \citenamefont {Lindner},
  \citenamefont {Fendley}, \citenamefont {Nayak}, \citenamefont {Oreg},
  \citenamefont {Stern}, \citenamefont {Berg}, \citenamefont {Shtengel},\ and\
  \citenamefont {Fisher}}]{mong2014}%
  \BibitemOpen
  \bibfield  {author} {\bibinfo {author} {\bibfnamefont {R.~S.~K.}\
  \bibnamefont {Mong}}, \bibinfo {author} {\bibfnamefont {D.~J.}\ \bibnamefont
  {Clarke}}, \bibinfo {author} {\bibfnamefont {J.}~\bibnamefont {Alicea}},
  \bibinfo {author} {\bibfnamefont {N.~H.}\ \bibnamefont {Lindner}}, \bibinfo
  {author} {\bibfnamefont {P.}~\bibnamefont {Fendley}}, \bibinfo {author}
  {\bibfnamefont {C.}~\bibnamefont {Nayak}}, \bibinfo {author} {\bibfnamefont
  {Y.}~\bibnamefont {Oreg}}, \bibinfo {author} {\bibfnamefont {A.}~\bibnamefont
  {Stern}}, \bibinfo {author} {\bibfnamefont {E.}~\bibnamefont {Berg}},
  \bibinfo {author} {\bibfnamefont {K.}~\bibnamefont {Shtengel}},\ and\
  \bibinfo {author} {\bibfnamefont {M.~P.~A.}\ \bibnamefont {Fisher}},\
  }\bibfield  {title} {\bibinfo {title} {Universal topological quantum
  computation from a superconductor-abelian quantum {H}all heterostructure},\
  }\href {https://doi.org/10.1103/PhysRevX.4.011036} {\bibfield  {journal}
  {\bibinfo  {journal} {Phys. Rev. X}\ }\textbf {\bibinfo {volume} {4}},\
  \bibinfo {pages} {011036} (\bibinfo {year} {2014})}\BibitemShut {NoStop}%
\bibitem [{\citenamefont {Clarke}\ \emph {et~al.}(2014)\citenamefont {Clarke},
  \citenamefont {Alicea},\ and\ \citenamefont {Shtengel}}]{clarke2014}%
  \BibitemOpen
  \bibfield  {author} {\bibinfo {author} {\bibfnamefont {D.~J.}\ \bibnamefont
  {Clarke}}, \bibinfo {author} {\bibfnamefont {J.}~\bibnamefont {Alicea}},\
  and\ \bibinfo {author} {\bibfnamefont {K.}~\bibnamefont {Shtengel}},\
  }\bibfield  {title} {\bibinfo {title} {Exotic circuit elements from
  zero-modes in hybrid superconductor--quantum-{H}all systems},\ }\href@noop {}
  {\bibfield  {journal} {\bibinfo  {journal} {Nature Physics}\ }\textbf
  {\bibinfo {volume} {10}},\ \bibinfo {pages} {877} (\bibinfo {year}
  {2014})}\BibitemShut {NoStop}%
\bibitem [{\citenamefont {Nayak}\ \emph {et~al.}(2008)\citenamefont {Nayak},
  \citenamefont {Simon}, \citenamefont {Stern}, \citenamefont {Freedman},\ and\
  \citenamefont {Das~Sarma}}]{nayak2008}%
  \BibitemOpen
  \bibfield  {author} {\bibinfo {author} {\bibfnamefont {C.}~\bibnamefont
  {Nayak}}, \bibinfo {author} {\bibfnamefont {S.~H.}\ \bibnamefont {Simon}},
  \bibinfo {author} {\bibfnamefont {A.}~\bibnamefont {Stern}}, \bibinfo
  {author} {\bibfnamefont {M.}~\bibnamefont {Freedman}},\ and\ \bibinfo
  {author} {\bibfnamefont {S.}~\bibnamefont {Das~Sarma}},\ }\bibfield  {title}
  {\bibinfo {title} {Non-abelian anyons and topological quantum computation},\
  }\href {https://doi.org/10.1103/RevModPhys.80.1083} {\bibfield  {journal}
  {\bibinfo  {journal} {Rev. Mod. Phys.}\ }\textbf {\bibinfo {volume} {80}},\
  \bibinfo {pages} {1083} (\bibinfo {year} {2008})}\BibitemShut {NoStop}%
\bibitem [{\citenamefont {Andreev}(1964)}]{andreev1964}%
  \BibitemOpen
  \bibfield  {author} {\bibinfo {author} {\bibfnamefont {A.}~\bibnamefont
  {Andreev}},\ }\bibfield  {title} {\bibinfo {title} {The thermal conductivity
  of the intermediate state in superconductors},\ }\href
  {http://jetp.ras.ru/cgi-bin/e/index/e/19/5/p1228?a=list} {\bibfield
  {journal} {\bibinfo  {journal} {Sov. Phys. JETP}\ }\textbf {\bibinfo {volume}
  {19}},\ \bibinfo {pages} {1228} (\bibinfo {year} {1964})}\BibitemShut
  {NoStop}%
\bibitem [{\citenamefont {Bozhko}\ \emph {et~al.}(1982)\citenamefont {Bozhko},
  \citenamefont {Tsoi},\ and\ \citenamefont {Yakovlev}}]{bozhko1982}%
  \BibitemOpen
  \bibfield  {author} {\bibinfo {author} {\bibfnamefont {S.}~\bibnamefont
  {Bozhko}}, \bibinfo {author} {\bibfnamefont {V.}~\bibnamefont {Tsoi}},\ and\
  \bibinfo {author} {\bibfnamefont {S.}~\bibnamefont {Yakovlev}},\ }\bibfield
  {title} {\bibinfo {title} {Observation of {A}ndreev reflection with the help
  of transverse electron focusing},\ }\href
  {http://jetpletters.ru/ps/1332/article_20128.shtml} {\bibfield  {journal}
  {\bibinfo  {journal} {JETP Letters}\ }\textbf {\bibinfo {volume} {36}},\
  \bibinfo {pages} {123} (\bibinfo {year} {1982})}\BibitemShut {NoStop}%
\bibitem [{\citenamefont {Benistant}\ \emph {et~al.}(1983)\citenamefont
  {Benistant}, \citenamefont {van Kempen},\ and\ \citenamefont
  {Wyder}}]{benistant1983}%
  \BibitemOpen
  \bibfield  {author} {\bibinfo {author} {\bibfnamefont {P.~A.~M.}\
  \bibnamefont {Benistant}}, \bibinfo {author} {\bibfnamefont {H.}~\bibnamefont
  {van Kempen}},\ and\ \bibinfo {author} {\bibfnamefont {P.}~\bibnamefont
  {Wyder}},\ }\bibfield  {title} {\bibinfo {title} {Direct observation of
  {A}ndreev reflection},\ }\href {https://doi.org/10.1103/PhysRevLett.51.817}
  {\bibfield  {journal} {\bibinfo  {journal} {Phys. Rev. Lett.}\ }\textbf
  {\bibinfo {volume} {51}},\ \bibinfo {pages} {817} (\bibinfo {year}
  {1983})}\BibitemShut {NoStop}%
\bibitem [{\citenamefont {Eroms}\ \emph {et~al.}(2005)\citenamefont {Eroms},
  \citenamefont {Weiss}, \citenamefont {Boeck}, \citenamefont {Borghs},\ and\
  \citenamefont {Z\"ulicke}}]{eroms05}%
  \BibitemOpen
  \bibfield  {author} {\bibinfo {author} {\bibfnamefont {J.}~\bibnamefont
  {Eroms}}, \bibinfo {author} {\bibfnamefont {D.}~\bibnamefont {Weiss}},
  \bibinfo {author} {\bibfnamefont {J.~D.}\ \bibnamefont {Boeck}}, \bibinfo
  {author} {\bibfnamefont {G.}~\bibnamefont {Borghs}},\ and\ \bibinfo {author}
  {\bibfnamefont {U.}~\bibnamefont {Z\"ulicke}},\ }\bibfield  {title} {\bibinfo
  {title} {{A}ndreev reflection at high magnetic fields: Evidence for electron
  and hole transport in edge states},\ }\href
  {https://doi.org/10.1103/PhysRevLett.95.107001} {\bibfield  {journal}
  {\bibinfo  {journal} {Phys. Rev. Lett.}\ }\textbf {\bibinfo {volume} {95}},\
  \bibinfo {pages} {107001} (\bibinfo {year} {2005})}\BibitemShut {NoStop}%
\bibitem [{\citenamefont {Batov}\ \emph {et~al.}(2007)\citenamefont {Batov},
  \citenamefont {Sch\"apers}, \citenamefont {Chtchelkatchev}, \citenamefont
  {Hardtdegen},\ and\ \citenamefont {Ustinov}}]{batov2007}%
  \BibitemOpen
  \bibfield  {author} {\bibinfo {author} {\bibfnamefont {I.~E.}\ \bibnamefont
  {Batov}}, \bibinfo {author} {\bibfnamefont {T.}~\bibnamefont {Sch\"apers}},
  \bibinfo {author} {\bibfnamefont {N.~M.}\ \bibnamefont {Chtchelkatchev}},
  \bibinfo {author} {\bibfnamefont {H.}~\bibnamefont {Hardtdegen}},\ and\
  \bibinfo {author} {\bibfnamefont {A.~V.}\ \bibnamefont {Ustinov}},\
  }\bibfield  {title} {\bibinfo {title} {Andreev reflection and strongly
  enhanced magnetoresistance oscillations in {G}a$_{x}${I}n$_{1-x}${As}/{I}n{P}
  heterostructures with superconducting contacts},\ }\href
  {https://doi.org/10.1103/PhysRevB.76.115313} {\bibfield  {journal} {\bibinfo
  {journal} {Phys. Rev. B}\ }\textbf {\bibinfo {volume} {76}},\ \bibinfo
  {pages} {115313} (\bibinfo {year} {2007})}\BibitemShut {NoStop}%
\bibitem [{\citenamefont {Nee}\ and\ \citenamefont {Prange}(1967)}]{nee1967}%
  \BibitemOpen
  \bibfield  {author} {\bibinfo {author} {\bibfnamefont {T.}~\bibnamefont
  {Nee}}\ and\ \bibinfo {author} {\bibfnamefont {R.}~\bibnamefont {Prange}},\
  }\bibfield  {title} {\bibinfo {title} {Quantum spectroscopy of the low field
  oscillations of the surface impedance},\ }\href
  {https://doi.org/https://doi.org/10.1016/0375-9601(67)90270-8} {\bibfield
  {journal} {\bibinfo  {journal} {Physics Lett. A}\ }\textbf {\bibinfo {volume}
  {25}},\ \bibinfo {pages} {583} (\bibinfo {year} {1967})}\BibitemShut
  {NoStop}%
\bibitem [{\citenamefont {Lee}\ \emph {et~al.}(2017)\citenamefont {Lee},
  \citenamefont {Huang}, \citenamefont {Efetov}, \citenamefont {Wei},
  \citenamefont {Hart}, \citenamefont {Taniguchi}, \citenamefont {Watanabe},
  \citenamefont {Yacoby},\ and\ \citenamefont {Kim}}]{lee2017}%
  \BibitemOpen
  \bibfield  {author} {\bibinfo {author} {\bibfnamefont {G.-H.}\ \bibnamefont
  {Lee}}, \bibinfo {author} {\bibfnamefont {K.-F.}\ \bibnamefont {Huang}},
  \bibinfo {author} {\bibfnamefont {D.~K.}\ \bibnamefont {Efetov}}, \bibinfo
  {author} {\bibfnamefont {D.~S.}\ \bibnamefont {Wei}}, \bibinfo {author}
  {\bibfnamefont {S.}~\bibnamefont {Hart}}, \bibinfo {author} {\bibfnamefont
  {T.}~\bibnamefont {Taniguchi}}, \bibinfo {author} {\bibfnamefont
  {K.}~\bibnamefont {Watanabe}}, \bibinfo {author} {\bibfnamefont
  {A.}~\bibnamefont {Yacoby}},\ and\ \bibinfo {author} {\bibfnamefont
  {P.}~\bibnamefont {Kim}},\ }\bibfield  {title} {\bibinfo {title} {Inducing
  superconducting correlation in quantum {H}all edge states},\ }\href@noop {}
  {\bibfield  {journal} {\bibinfo  {journal} {Nature Physics}\ }\textbf
  {\bibinfo {volume} {13}},\ \bibinfo {pages} {693} (\bibinfo {year}
  {2017})}\BibitemShut {NoStop}%
\bibitem [{\citenamefont {Zhao}\ \emph {et~al.}(2020)\citenamefont {Zhao},
  \citenamefont {Arnault}, \citenamefont {Bondarev}, \citenamefont
  {Seredinski}, \citenamefont {Larson}, \citenamefont {Draelos}, \citenamefont
  {Li}, \citenamefont {Watanabe}, \citenamefont {Taniguchi}, \citenamefont
  {Amet}, \citenamefont {Baranger},\ and\ \citenamefont
  {Finkelstein}}]{zhao2020}%
  \BibitemOpen
  \bibfield  {author} {\bibinfo {author} {\bibfnamefont {L.}~\bibnamefont
  {Zhao}}, \bibinfo {author} {\bibfnamefont {E.~G.}\ \bibnamefont {Arnault}},
  \bibinfo {author} {\bibfnamefont {A.}~\bibnamefont {Bondarev}}, \bibinfo
  {author} {\bibfnamefont {A.}~\bibnamefont {Seredinski}}, \bibinfo {author}
  {\bibfnamefont {T.~F.~Q.}\ \bibnamefont {Larson}}, \bibinfo {author}
  {\bibfnamefont {A.~W.}\ \bibnamefont {Draelos}}, \bibinfo {author}
  {\bibfnamefont {H.}~\bibnamefont {Li}}, \bibinfo {author} {\bibfnamefont
  {K.}~\bibnamefont {Watanabe}}, \bibinfo {author} {\bibfnamefont
  {T.}~\bibnamefont {Taniguchi}}, \bibinfo {author} {\bibfnamefont
  {F.}~\bibnamefont {Amet}}, \bibinfo {author} {\bibfnamefont {H.~U.}\
  \bibnamefont {Baranger}},\ and\ \bibinfo {author} {\bibfnamefont
  {G.}~\bibnamefont {Finkelstein}},\ }\bibfield  {title} {\bibinfo {title}
  {Interference of chiral {A}ndreev edge states},\ }\href@noop {} {\bibfield
  {journal} {\bibinfo  {journal} {Nature Physics}\ }\textbf {\bibinfo {volume}
  {16}},\ \bibinfo {pages} {862} (\bibinfo {year} {2020})}\BibitemShut
  {NoStop}%
\bibitem [{\citenamefont {G\"ul}\ \emph {et~al.}(2021)\citenamefont {G\"ul},
  \citenamefont {Ronen}, \citenamefont {Lee}, \citenamefont {Shapourian},
  \citenamefont {Zauberman}, \citenamefont {Lee}, \citenamefont {Watanabe},
  \citenamefont {Taniguchi}, \citenamefont {Vishwanath}, \citenamefont
  {Yacoby},\ and\ \citenamefont {Kim}}]{gul2021}%
  \BibitemOpen
  \bibfield  {author} {\bibinfo {author} {\bibfnamefont {O.}~\bibnamefont
  {G\"ul}}, \bibinfo {author} {\bibfnamefont {Y.}~\bibnamefont {Ronen}},
  \bibinfo {author} {\bibfnamefont {S.~Y.}\ \bibnamefont {Lee}}, \bibinfo
  {author} {\bibfnamefont {H.}~\bibnamefont {Shapourian}}, \bibinfo {author}
  {\bibfnamefont {J.}~\bibnamefont {Zauberman}}, \bibinfo {author}
  {\bibfnamefont {Y.~H.}\ \bibnamefont {Lee}}, \bibinfo {author} {\bibfnamefont
  {K.}~\bibnamefont {Watanabe}}, \bibinfo {author} {\bibfnamefont
  {T.}~\bibnamefont {Taniguchi}}, \bibinfo {author} {\bibfnamefont
  {A.}~\bibnamefont {Vishwanath}}, \bibinfo {author} {\bibfnamefont
  {A.}~\bibnamefont {Yacoby}},\ and\ \bibinfo {author} {\bibfnamefont
  {P.}~\bibnamefont {Kim}},\ }\href@noop {} {\bibinfo {title} {Andreev
  reflection in the fractional quantum {H}all state}} (\bibinfo {year}
  {2021}),\ \Eprint {https://arxiv.org/abs/2009.07836} {arXiv:2009.07836
  [cond-mat.mes-hall]} \BibitemShut {NoStop}%
\bibitem [{\citenamefont {Hatefipour}\ \emph {et~al.}(2021)\citenamefont
  {Hatefipour}, \citenamefont {Cuozzo}, \citenamefont {Kanter}, \citenamefont
  {Strickland}, \citenamefont {Lu}, \citenamefont {Rossi},\ and\ \citenamefont
  {Shabani}}]{hatefipour2021}%
  \BibitemOpen
  \bibfield  {author} {\bibinfo {author} {\bibfnamefont {M.}~\bibnamefont
  {Hatefipour}}, \bibinfo {author} {\bibfnamefont {J.~J.}\ \bibnamefont
  {Cuozzo}}, \bibinfo {author} {\bibfnamefont {J.}~\bibnamefont {Kanter}},
  \bibinfo {author} {\bibfnamefont {W.}~\bibnamefont {Strickland}}, \bibinfo
  {author} {\bibfnamefont {T.-M.}\ \bibnamefont {Lu}}, \bibinfo {author}
  {\bibfnamefont {E.}~\bibnamefont {Rossi}},\ and\ \bibinfo {author}
  {\bibfnamefont {J.}~\bibnamefont {Shabani}},\ }\href@noop {} {\bibinfo
  {title} {Induced superconducting pairing in integer quantum {H}all edge
  states}} (\bibinfo {year} {2021}),\ \Eprint
  {https://arxiv.org/abs/2108.08899} {arXiv:2108.08899 [cond-mat.mes-hall]}
  \BibitemShut {NoStop}%
\bibitem [{\citenamefont {Manesco}\ \emph {et~al.}(2021)\citenamefont
  {Manesco}, \citenamefont {Flór}, \citenamefont {Liu},\ and\ \citenamefont
  {Akhmerov}}]{manesco21}%
  \BibitemOpen
  \bibfield  {author} {\bibinfo {author} {\bibfnamefont {A.~L.~R.}\
  \bibnamefont {Manesco}}, \bibinfo {author} {\bibfnamefont {I.~M.}\
  \bibnamefont {Flór}}, \bibinfo {author} {\bibfnamefont {C.-X.}\ \bibnamefont
  {Liu}},\ and\ \bibinfo {author} {\bibfnamefont {A.~R.}\ \bibnamefont
  {Akhmerov}},\ }\href@noop {} {\bibinfo {title} {Mechanisms of {A}ndreev
  reflection in quantum {Hall} graphene}} (\bibinfo {year} {2021}),\ \Eprint
  {https://arxiv.org/abs/2103.06722} {arXiv:2103.06722 [cond-mat.mes-hall]}
  \BibitemShut {NoStop}%
\bibitem [{\citenamefont {Altshuler}(1985)}]{altshuler1985}%
  \BibitemOpen
  \bibfield  {author} {\bibinfo {author} {\bibfnamefont {B.~L.}\ \bibnamefont
  {Altshuler}},\ }\bibfield  {title} {\bibinfo {title} {Fluctuations in the
  extrinsic conductivity of disordered conductors},\ }\href
  {http://jetpletters.ru/ps/1470/article_22425.shtml} {\bibfield  {journal}
  {\bibinfo  {journal} {JETP Letters}\ }\textbf {\bibinfo {volume} {41}},\
  \bibinfo {pages} {530} (\bibinfo {year} {1985})}\BibitemShut {NoStop}%
\bibitem [{\citenamefont {Lee}\ and\ \citenamefont {Stone}(1985)}]{lee1985}%
  \BibitemOpen
  \bibfield  {author} {\bibinfo {author} {\bibfnamefont {P.~A.}\ \bibnamefont
  {Lee}}\ and\ \bibinfo {author} {\bibfnamefont {A.~D.}\ \bibnamefont
  {Stone}},\ }\bibfield  {title} {\bibinfo {title} {Universal conductance
  fluctuations in metals},\ }\href
  {https://doi.org/10.1103/PhysRevLett.55.1622} {\bibfield  {journal} {\bibinfo
   {journal} {Phys. Rev. Lett.}\ }\textbf {\bibinfo {volume} {55}},\ \bibinfo
  {pages} {1622} (\bibinfo {year} {1985})}\BibitemShut {NoStop}%
\bibitem [{\citenamefont {Prada}\ and\ \citenamefont {Sols}(2004)}]{prada2004}%
  \BibitemOpen
  \bibfield  {author} {\bibinfo {author} {\bibfnamefont {E.}~\bibnamefont
  {Prada}}\ and\ \bibinfo {author} {\bibfnamefont {F.}~\bibnamefont {Sols}},\
  }\bibfield  {title} {\bibinfo {title} {Entangled electron current through
  finite size normal-superconductor tunneling structures},\ }\href@noop {}
  {\bibfield  {journal} {\bibinfo  {journal} {The European Physical Journal B -
  Condensed Matter and Complex Systems}\ }\textbf {\bibinfo {volume} {40}},\
  \bibinfo {pages} {379} (\bibinfo {year} {2004})}\BibitemShut {NoStop}%
\bibitem [{\citenamefont {Lutchyn}\ \emph {et~al.}(2012)\citenamefont
  {Lutchyn}, \citenamefont {Stanescu},\ and\ \citenamefont
  {Das~Sarma}}]{lutchyn2012}%
  \BibitemOpen
  \bibfield  {author} {\bibinfo {author} {\bibfnamefont {R.~M.}\ \bibnamefont
  {Lutchyn}}, \bibinfo {author} {\bibfnamefont {T.~D.}\ \bibnamefont
  {Stanescu}},\ and\ \bibinfo {author} {\bibfnamefont {S.}~\bibnamefont
  {Das~Sarma}},\ }\bibfield  {title} {\bibinfo {title} {Momentum relaxation in
  a semiconductor proximity-coupled to a disordered $s$-wave superconductor:
  Effect of scattering on topological superconductivity},\ }\href
  {https://doi.org/10.1103/PhysRevB.85.140513} {\bibfield  {journal} {\bibinfo
  {journal} {Phys. Rev. B}\ }\textbf {\bibinfo {volume} {85}},\ \bibinfo
  {pages} {140513} (\bibinfo {year} {2012})}\BibitemShut {NoStop}%
\bibitem [{\citenamefont {Hekking}\ and\ \citenamefont
  {Nazarov}(1994)}]{hekking1994}%
  \BibitemOpen
  \bibfield  {author} {\bibinfo {author} {\bibfnamefont {F.~W.~J.}\
  \bibnamefont {Hekking}}\ and\ \bibinfo {author} {\bibfnamefont {Y.~V.}\
  \bibnamefont {Nazarov}},\ }\bibfield  {title} {\bibinfo {title} {Subgap
  conductivity of a superconductor--normal-metal tunnel interface},\ }\href
  {https://doi.org/10.1103/PhysRevB.49.6847} {\bibfield  {journal} {\bibinfo
  {journal} {Phys. Rev. B}\ }\textbf {\bibinfo {volume} {49}},\ \bibinfo
  {pages} {6847} (\bibinfo {year} {1994})}\BibitemShut {NoStop}%
\bibitem [{sm()}]{sm}%
  \BibitemOpen
  \href@noop {} {\bibinfo {title} {See {S}upplemental {M}aterial for
  details.}}\BibitemShut {Stop}%
\bibitem [{Note1()}]{Note1}%
  \BibitemOpen
  \bibinfo {note} {In the parameterization of $a_{\protect \rm e}$ and
  $a_{\protect \rm h}$, we suppressed the common phase which is inconsequential
  for $G$.}\BibitemShut {Stop}%
\bibitem [{\citenamefont {Van~Kampen}(2007)}]{vankampen2007}%
  \BibitemOpen
  \bibfield  {author} {\bibinfo {author} {\bibfnamefont {N.}~\bibnamefont
  {Van~Kampen}},\ }\href
  {https://doi.org/https://doi.org/10.1016/B978-0-444-52965-7.X5000-4} {\emph
  {\bibinfo {title} {Stochastic Processes in Physics and Chemistry}}},\
  \bibinfo {edition} {3rd}\ ed.\ (\bibinfo  {publisher} {North Holland},\
  \bibinfo {address} {Amsterdam},\ \bibinfo {year} {2007})\BibitemShut
  {NoStop}%
\bibitem [{Note2()}]{Note2}%
  \BibitemOpen
  \bibinfo {note} {Exponential decay of $\langle G \rangle $ with $L$ was also
  noted in Ref.~\cite {zhang2020}.}\BibitemShut {Stop}%
\bibitem [{\citenamefont {Watts-Tobin}\ and\ \citenamefont
  {Waterworth}(1973)}]{wattstobin1973}%
  \BibitemOpen
  \bibfield  {author} {\bibinfo {author} {\bibfnamefont {R.~J.}\ \bibnamefont
  {Watts-Tobin}}\ and\ \bibinfo {author} {\bibfnamefont {G.~M.}\ \bibnamefont
  {Waterworth}},\ }\bibfield  {title} {\bibinfo {title} {Calculation of the
  vortex structure in a superconducting alloy},\ }\href
  {https://doi.org/10.1007/BF01391916} {\bibfield  {journal} {\bibinfo
  {journal} {Zeitschrift f\"ur Physik A Hadrons and nuclei}\ }\textbf {\bibinfo
  {volume} {261}},\ \bibinfo {pages} {249} (\bibinfo {year}
  {1973})}\BibitemShut {NoStop}%
\bibitem [{\citenamefont {Lian}\ \emph {et~al.}(2016)\citenamefont {Lian},
  \citenamefont {Wang},\ and\ \citenamefont {Zhang}}]{lian2016}%
  \BibitemOpen
  \bibfield  {author} {\bibinfo {author} {\bibfnamefont {B.}~\bibnamefont
  {Lian}}, \bibinfo {author} {\bibfnamefont {J.}~\bibnamefont {Wang}},\ and\
  \bibinfo {author} {\bibfnamefont {S.-C.}\ \bibnamefont {Zhang}},\ }\bibfield
  {title} {\bibinfo {title} {Edge-state-induced {A}ndreev oscillation in
  quantum anomalous {H}all insulator-superconductor junctions},\ }\href
  {https://doi.org/10.1103/PhysRevB.93.161401} {\bibfield  {journal} {\bibinfo
  {journal} {Phys. Rev. B}\ }\textbf {\bibinfo {volume} {93}},\ \bibinfo
  {pages} {161401} (\bibinfo {year} {2016})}\BibitemShut {NoStop}%
\bibitem [{\citenamefont {Slichter}(1990)}]{slichter1990}%
  \BibitemOpen
  \bibfield  {author} {\bibinfo {author} {\bibfnamefont {C.~P.}\ \bibnamefont
  {Slichter}},\ }\href@noop {} {\emph {\bibinfo {title} {Principles of Magnetic
  Resonance}}},\ \bibinfo {edition} {3rd}\ ed.\ (\bibinfo  {publisher}
  {Springer-Verlag},\ \bibinfo {address} {Berlin},\ \bibinfo {year}
  {1990})\BibitemShut {NoStop}%
\bibitem [{Note3()}]{Note3}%
  \BibitemOpen
  \bibinfo {note} {The high compressibility means small disorder-induced
  broadening of Landau levels, $\delta \varepsilon \ll \hbar \omega _{\protect
  \rm c}$. We also assumed the London penetration depth $\lambda \gg
  l_{\protect \rm A}$ to neglect the diamagnetic current effect on $\delta
  k_\mu $.}\BibitemShut {Stop}%
\bibitem [{Note4()}]{Note4}%
  \BibitemOpen
  \bibinfo {note} {We assume $\Delta \ll \hbar \omega _{\protect \rm c}$ and
  disregard other mechanisms controlled by $E / (\hbar \omega _{\protect \rm
  c})$.}\BibitemShut {Stop}%
\bibitem [{\citenamefont {Imambekov}\ \emph {et~al.}(2012)\citenamefont
  {Imambekov}, \citenamefont {Schmidt},\ and\ \citenamefont
  {Glazman}}]{imambekov2012}%
  \BibitemOpen
  \bibfield  {author} {\bibinfo {author} {\bibfnamefont {A.}~\bibnamefont
  {Imambekov}}, \bibinfo {author} {\bibfnamefont {T.~L.}\ \bibnamefont
  {Schmidt}},\ and\ \bibinfo {author} {\bibfnamefont {L.~I.}\ \bibnamefont
  {Glazman}},\ }\bibfield  {title} {\bibinfo {title} {One-dimensional quantum
  liquids: Beyond the {L}uttinger liquid paradigm},\ }\href
  {https://doi.org/10.1103/RevModPhys.84.1253} {\bibfield  {journal} {\bibinfo
  {journal} {Rev. Mod. Phys.}\ }\textbf {\bibinfo {volume} {84}},\ \bibinfo
  {pages} {1253} (\bibinfo {year} {2012})}\BibitemShut {NoStop}%
\bibitem [{\citenamefont {Kane}\ and\ \citenamefont {Fisher}(1995)}]{kane1995}%
  \BibitemOpen
  \bibfield  {author} {\bibinfo {author} {\bibfnamefont {C.~L.}\ \bibnamefont
  {Kane}}\ and\ \bibinfo {author} {\bibfnamefont {M.~P.~A.}\ \bibnamefont
  {Fisher}},\ }\bibfield  {title} {\bibinfo {title} {Impurity scattering and
  transport of fractional quantum {H}all edge states},\ }\href
  {https://doi.org/10.1103/PhysRevB.51.13449} {\bibfield  {journal} {\bibinfo
  {journal} {Phys. Rev. B}\ }\textbf {\bibinfo {volume} {51}},\ \bibinfo
  {pages} {13449} (\bibinfo {year} {1995})}\BibitemShut {NoStop}%
\bibitem [{\citenamefont {Zhang}\ and\ \citenamefont {Liu}(2020)}]{zhang2020}%
  \BibitemOpen
  \bibfield  {author} {\bibinfo {author} {\bibfnamefont {J.-X.}\ \bibnamefont
  {Zhang}}\ and\ \bibinfo {author} {\bibfnamefont {C.-X.}\ \bibnamefont
  {Liu}},\ }\bibfield  {title} {\bibinfo {title} {Disordered quantum transport
  in quantum anomalous {Hall} insulator-superconductor junctions},\ }\href
  {https://doi.org/10.1103/PhysRevB.102.144513} {\bibfield  {journal} {\bibinfo
   {journal} {Physical Review B}\ }\textbf {\bibinfo {volume} {102}},\ \bibinfo
  {pages} {144513} (\bibinfo {year} {2020})}\BibitemShut {NoStop}%
\end{thebibliography}%

\end{document}

\title{Supplemental Material for ``Disorder in Andreev reflection of a quantum Hall edge''}

\author{Vladislav D.~Kurilovich}
\affiliation{Department of Physics, Yale University, New Haven, CT 06520, USA}
\author{Zachary M.~Raines}
\affiliation{Department of Physics, Yale University, New Haven, CT 06520, USA}
\author{Leonid I.~Glazman}
\affiliation{Department of Physics, Yale University, New Haven, CT 06520, USA}

\maketitle

\onecolumngrid
\vspace*{-5mm}

\section{Derivation of $1/\lA$, Eq.~(8)\label{sec:lA}}

In this section, we present details of the derivation of Eq.~(8).
We start with the obtained in the main text expression for the Andreev amplitude, see Eq.~(7). For convenience, we reproduce it here:
\begin{align}\label{eq:amplitudes}
    \he &= -\frac{(\partial_y \Phi)^2 t^2}{v}\int_0^\lprox dx_1 dx_2 e^{i\kf (x_1 + x_2)}\partial^2_{y_1 y_2}{\cal G}_{\rm he} (x_1, x_2),
\end{align}
(we remind that ${\cal G}_{\rm he}(x_1, x_2) \equiv {\cal G}_{\rm he}(\bm r_1, \bm r_2 | E = 0)|_{y_{1,2}, z_{1,2} = 0}$ is the anomalous Green's function of the superconductor).
For calculations, it is convenient to choose a gauge in which the vector potential vanishes at the interface between the superconductor and the 2DEG.
In this gauge, the wave vector $k_\mu$ in Eq.~\eqref{eq:amplitudes} is related to the distance $y_{\rm c}$ between the cyclotron orbit center and the interface: $k_\mu = y_{\rm c} / l_B^2$, where $l_B = \sqrt{c / e B}$ is the magnetic length (throughout the supplement, we work in units with $\hbar = 1$). At $\nu = 2$, we can thus estimate 
\begin{equation}
    k_\mu \lesssim 1 / l_B.
\end{equation}

For simplicity, we first consider a type I superconductor. In this case, the vector potential vanishes not only at the interface but everywhere in the superconductor. We describe the superconductor with the standard BCS Hamiltonian:
\begin{equation}\label{eq:H_SC}
    H_{\rm SC} = \sum_\sigma\int d^3r\,\chi^\dagger_\sigma({\bm r})\Big[-\frac{\partial_{\bm r}^2}{2m} - \mu + U({\bm r})\Big] \chi_\sigma({\bm r}) + \int d^3 r\,\Delta \big(\chi_\uparrow^\dagger ({\bm r})\chi_\downarrow^\dagger({\bm r}) + \chi_\downarrow ({\bm r})\chi_\uparrow ({\bm r})\big).
\end{equation}
Here $\chi_\sigma(\bm r)$ is an annihilation operator for an electron with spin $\sigma$, $m$ is the effective mass, $\mu$ is the chemical potential, and $\Delta$ is the superconducting order parameter. 
$U({\bm r})$ is the disorder potential, which we assume to be a Gaussian random variable with a short-ranged correlation function,
\begin{equation}\label{eq:disorder}
    \langle U({\bm r}) U({\bm r}^\prime)\rangle = \frac{1}{2\pi \nu_{\rm M} \tau_{\rm mfp}} \delta({\bm r} - {\bm r}^\prime).
\end{equation}
We parameterized the correlation function by the normal-state density of states in the metal $\nu_{\rm M}$ and the electron mean free time $\tau_{\rm mfp}$. We assume that the superconductor is ``dirty'',  $\Delta\cdot \tau_{\rm mfp} \ll 1$. 

Let us now compute the average probability of the Andreev reflection (our approach is similar in spirit to that in Ref.~\onlinecite{hek94}).
Using Eq.~\eqref{eq:amplitudes}, we first represent $\langle |\he|^2 \rangle$ as
\begin{align}\label{eq:var}
    \langle |\he |^2 \rangle = \frac{(\partial_y \Phi)^4 t^4}{v^2}\int^{\lprox}_{0} dx_{1}dx_{2}dx_{3}dx_{4}e^{ik_\mu(x_1 +x_2)}e^{-ik_\mu(x_3 + x_4)}\partial^2_{y_1,y_2}\partial^2_{y_3, y_4}\Lngl {\cal G}_{\rm he}(\bm r_1, \bm r_2|0)\cdot {\cal G}_{\rm eh}(\bm r_4, \bm r_3|0)\Rngl\big|_{y_\alpha, z_\alpha = 0}.
\end{align}
On the right hand side, we replaced the average by its irreducible component. This is possible because $\langle \he \rangle = 0$ (see discussion in the main text). 
The superconductor Green's functions in Eq.~\eqref{eq:var} can be expressed in terms of the retarded Green's function $\GR$ of the metal in the normal state:
\begin{equation}\label{eq:greeen}
    {\cal G} ({\bm r}_1, {\bm r}_2| E) = \int \frac{d\epsilon}{\Delta^2 - E^2 + \epsilon^2} \begin{pmatrix}
    E+\epsilon & \Delta\\
    \Delta & E - \epsilon
    \end{pmatrix} \frac{1}{\pi}\,{\rm Im}\,\GR ({\bm r}_1, {\bm r}_2| \epsilon).
\end{equation}
Substituting this relation with $E = 0$ into Eq.~\eqref{eq:var} we obtain
\begin{align}\label{eq:var2}
   \langle |\he|^2 \rangle  =&  \frac{(\partial_y \Phi)^4 t^4}{\pi^2 v^2}\int^{\lprox}_{0} dx_{1}dx_{2}dx_{3}dx_{4}e^{ik_\mu(x_1 +x_2)}e^{-ik_\mu(x_3 + x_4)}\int \frac{\Delta d\epsilon}{\Delta^2 +\epsilon^2} \frac{\Delta d\epsilon^\prime}{\Delta^2+\epsilon^{\prime 2}}\notag\\
    &\times\partial^2_{y_1,y_2}\partial^2_{y_3, y_4}\Lngl {\rm Im}\, \GR(\bm r_1, \bm r_2|\epsilon)\cdot{\rm Im}\, \GR(\bm r_4, \bm r_3|\epsilon^\prime)\Rngl\big|_{y_\alpha, z_\alpha = 0}.
\end{align}
Let us focus on the averaged-over-disorder product of the Green's functions in the above expression. We can represent this product as
\begin{align}\label{eq:two_conts}
    \Lngl {\rm Im}\, \GR({\bm r}_1,{\bm r}_2|\epsilon)&\cdot {\rm Im}\, \GR({\bm r}_4, {\bm r}_3|\epsilon^\prime)\Rngl = \frac{1}{2}{\rm Re}\big[\Lngl \GR({\bm r}_1,{\bm r}_2|\epsilon)\cdot \GA({\bm r}_4, {\bm r}_3|\epsilon^\prime)\Rngl - \Lngl \GR({\bm r}_1,{\bm r}_2|\epsilon)\cdot \GR({\bm r}_4, {\bm r}_3|\epsilon^\prime)\Rngl\big],
\end{align}
where $\GA$ is the advanced normal state Green's function. We will see below that the contribution of the first term to $\langle |\he|^2\rangle$ is determined by long diffusive electron trajectories of size $\sim \xic$ ($\xic$ is the superconducting coherence length). On the other hand, the contribution of the second term is determined by trajectories of length $\lesssim \lambda_F$ only ($\lambda_F$ is the Fermi wave length in the superconductor). This means that the latter contribution is small compared to the one produced by the first term in Eq.~\eqref{eq:two_conts}. In what follows we neglect the second term.

The average $\lngl \GR \cdot \GA\rngl$ can be expressed in terms of the normal-state diffuson and Cooperon \cite{aleiner96}. 
Using Eq.~\eqref{eq:disorder} and neglecting small corrections that have a relative magnitude $\sim \lambda_F / l_{\rm mfp} \ll 1$ (with $l_{\rm mfp} = v_F \tau_{\rm mfp}$ being the mean free path), we represent $\lngl \GR \cdot \GA\rngl$ as
\begin{align}
    \Lngl \GR({\bm r}_1,{\bm r}_2|\epsilon)&\cdot \GA({\bm r}_4, {\bm r}_3|\epsilon^\prime)\Rngl\notag\\
    &= \frac{1}{2\pi\nu_{\rm M}\tau_{\rm mfp}^{2}}\int d^3rd^3r^\prime\,\langle \GR({\bm r}_{1},{\bm r}|\epsilon)\rangle\langle \GA({\bm r},{\bm r}_{3}|\epsilon^{\prime})\rangle \difz ({\bm r},{\bm r}^\prime|\epsilon-\epsilon^{\prime})\langle \GA({\bm r}_4,{\bm r}^\prime|\epsilon^{\prime})\rangle\langle \GR({\bm r}^\prime,{\bm r}_{2}|\epsilon)\rangle\label{eq:diffuson}\\
    &+ \frac{1}{2\pi\nu_{\rm M}\tau_{\rm mfp}^{2}}\int d^3rd^3r^\prime\,\langle \GR({\bm r}_{1},{\bm r}|\epsilon)\rangle\langle \GA({\bm r},{\bm r}_{4}|\epsilon^{\prime})\rangle \coop ({\bm r},{\bm r}^\prime|\epsilon-\epsilon^{\prime})\langle \GA({\bm r}_{3},{\bm r}^\prime|\epsilon^{\prime})\rangle\langle \GR({\bm r}^\prime,{\bm r}_{2} |\epsilon)\rangle.\label{eq:cooperon}
\end{align}
Here functions ${\difz}({\bm r},{\bm r}^\prime|\epsilon-\epsilon^{\prime})$ and ${\coop}({\bm r},{\bm r}^\prime|\epsilon-\epsilon^{\prime})$ are the diffuson and the Cooperon, respectively. 
The magnetic field does not penetrate a type I superconductor, so $\difz({\bm r},{\bm r}^\prime|\epsilon-\epsilon^{\prime}) = \coop({\bm r},{\bm r}^\prime|\epsilon-\epsilon^{\prime})$ in this case. In the time domain, ${\difz}({\bm r},{\bm r}^\prime|t)$ satisfies the diffusion equation \cite{aleiner96},
\begin{equation}\label{eq:diffusion}
    (\partial_t - D \partial_{\bm r}^2) {\difz}({\bm r},{\bm r}^\prime|t) = \delta(t)\delta({\bm r} - {\bm r}^\prime),
\end{equation}
with the boundary condition corresponding to the vanishing of the probability current at the metal's surface. Here $D = v_F l_{\rm mfp} / 3$ is the diffusion constant.

At relevant energies $\epsilon - \epsilon^\prime \sim \Delta$, the diffuson ${\difz}({\bm r},{\bm r}^\prime|\epsilon - \epsilon^\prime)$ varies at a length scale of the order of $\xic$. The latter satisfies $\xic \gg l_{\rm mfp}$ for a dirty superconductor. At the same time, the average Green's functions decay at a distance $\sim l_{\rm mfp}$. This means that in Eqs.~\eqref{eq:diffuson} and \eqref{eq:cooperon} the argument ${\bm r}$ of $\difz$ and $\coop$ is close to ${\bm r}_1$ and the argument ${\bm r}^\prime$ is close to ${\bm r}_2$. Consequently, we can approximate $\lngl \GR \cdot \GA \rngl$ as
\begin{align}\label{eq:RA}
    \Lngl \GR ({\bm r}_1,{\bm r}_2|\epsilon)\cdot \GA({\bm r}_4, {\bm r}_3|\epsilon^\prime)\Rngl
    = 2\pi \nu_{\rm M}\,{\difz}({\bm r}_1, {\bm r}_2|\epsilon - \epsilon^\prime) [V({\bm r}_1, {\bm r}_3)V({\bm r}_2, {\bm r}_4) + V({\bm r}_1, {\bm r}_4)V({\bm r}_2, {\bm r}_3)], 
\end{align}
where we abbreviated
\begin{equation}\label{eq:vertex}
    V({\bm r}_1, {\bm r}_3) = \frac{1}{2\pi\nu_{\rm M} \tau_{\rm mfp}}\int d^3r\,\langle \GR({\bm r}_{1},{\bm r}|\epsilon)\rangle\langle \GA({\bm r},{\bm r}_3|\epsilon^{\prime})\rangle.
\end{equation}
Combining Eqs.~\eqref{eq:var2}, \eqref{eq:two_conts}, and \eqref{eq:RA}, we obtain the following expression for $\langle |\he|^2 \rangle$:
\begin{align}\label{eq:var3}
   \langle |\he|^2 \rangle = & \frac{\nu_{\rm M}(\partial_y \Phi)^4 t^4}{\pi v^2}\int^{\lprox}_{0} dx_{1}dx_{2}dx_{3}dx_{4}e^{ik_\mu(x_1 +x_2)}e^{-ik_\mu(x_3 + x_4)}\int \frac{\Delta d\epsilon}{\Delta^2 +\epsilon^2} \frac{\Delta d\epsilon^\prime}{\Delta^2+\epsilon^{\prime 2}} \notag\\
    &\times{\rm Re}\,{\difz}(x_1,x_2|\epsilon - \epsilon^\prime)\,\partial^2_{y_1,y_2}\partial^2_{y_3, y_4}[V({\bm r}_1, {\bm r}_3)V({\bm r}_2, {\bm r}_4) + V({\bm r}_1, {\bm r}_4)V({\bm r}_2, {\bm r}_3)]\big|_{y_\alpha, z_\alpha = 0},
\end{align}
where ${\difz}(x_1,x_2|\epsilon - \epsilon^\prime) \equiv {\difz}(\bm r_1,\bm r_2|\epsilon - \epsilon^\prime)|_{y_{1,2},z_{1,2} = 0}$.

So far, we have been focusing on the case of a type I superconductor. 
Type II superconductor is different in that it admits magnetic field $B$.
The field affects functions ${\difz}$ and ${\coop}$ leading to additional phase factors in them. At relevant distances $\sim \xic$ the corresponding phases can be estimated as $\sim B \xic^2 / \phi_0 \sim B / H_{\rm c2}$ ($\phi_0$ is the flux quantum and $H_{\rm c2}$ is the upper critical field).
We see that for fields $B \ll H_{\rm c2}$ the phases are small and can be disregarded.
This means the derived at $B = 0$ Eq.~\eqref{eq:var3} is also applicable for the case of a type II superconductor in the regime $B \ll H_{\rm c2}$.
The same holds for all of the results presented in the remainder of the section.

Let us proceed with the derivation of $1/\lA$. Functions $V$ in Eq.~\eqref{eq:var3} stipulate ${\bm r}_1 \approx {\bm r}_3$, ${\bm r}_2 \approx {\bm r}_4$ in the diffuson's contribution and ${\bm r}_1 \approx {\bm r}_4$, ${\bm r}_2 \approx {\bm r}_3$ in the Cooperon's contribution. By making a direct calculation of the integral in Eq.~\eqref{eq:vertex}, we find for the combination of functions $V$ in Eq.~\eqref{eq:var3}:
\begin{align}\label{eq:vertex_expr}
    \partial^2_{y_1,y_2}\partial^2_{y_3, y_4}[V({\bm r}_1, {\bm r}_3)V({\bm r}_2, {\bm r}_4) + & V({\bm r}_1, {\bm r}_4)V({\bm r}_2, {\bm r}_3)]\big|_{y_\alpha, z_\alpha = 0}\notag\\
    &= (\pi p_F)^2 \big[\delta(x_1 -x_3) \delta(x_2-x_4) + \delta(x_1 -x_4)\delta(x_2-x_3)\big],
\end{align}
where $p_F$ is the Fermi momentum of the superconductor. The delta-functions in this expression should be interpreted as peaks of width $\sim \lambda_F$. With the help of Eq.~\eqref{eq:vertex_expr}, we can rewrite Eq.~\eqref{eq:var3} as
\begin{align}\label{eq:var4}
   \langle |\he|^2 \rangle = \frac{2\pi \nu_{\rm M}(\partial_y \Phi)^4 t^4 p_F^2}{v^2} \int_0^\lprox dx_{1}dx_{2}\int \frac{\Delta d\epsilon}{\Delta^2 +\epsilon^2} \frac{\Delta d\epsilon^\prime}{\Delta^2+\epsilon^{\prime 2}}
   {\rm Re}\,{\difz}(x_1,x_2|\epsilon - \epsilon^\prime).
\end{align}
The expression for ${\difz}(x_1,x_2|\epsilon - \epsilon^\prime)$ is sensitive to a particular geometry of the considered device. We will assume that the width of the superconducting film exceeds $\xic$. In this case, the film can be regarded as being three-dimensional for diffusion. We then find:
\begin{equation}\label{eq:diffuson_expl}
    {\difz}(x_1,x_2|\epsilon - \epsilon^\prime) = 2 \int_0^{+\infty}\frac{dt }{(4\pi D t)^{3/2}}e^{-i(\epsilon - \epsilon^\prime)t - \frac{(x_1 - x_2)^2}{4Dt}}
\end{equation}
(the factor of $2$ results from the boundary condition for Eq.~\eqref{eq:diffusion}).
Using this expression, one can easily show that
\begin{equation}\label{eq:diff_int}
    \int \frac{\Delta d\epsilon}{\Delta^2 +\epsilon^2} \frac{\Delta d\epsilon^\prime}{\Delta^2+\epsilon^{\prime 2}}
   {\rm Re}\,{\difz}(x_1,x_2|\epsilon - \epsilon^\prime) = \frac{\pi}{2D|x_{1}-x_{2}|}e^{-|x_{1}-x_{2}|/\xic},\quad\quad \xic = \sqrt{\frac{D}{2\Delta}}.
\end{equation}
We will assume that the length of the proximitized segment exceeds the coherence length, $\lprox \gg \xic$. Then, using Eq.~\eqref{eq:diff_int} in Eq.~\eqref{eq:var4} we obtain
\begin{equation}\label{eq:finalish}
   \langle |\he|^2 \rangle = \frac{2\pi^2 \nu_{\rm M}(\partial_y \Phi)^4 t^4 p_F^2}{v^2 D}
   \int_0^{+\infty} \frac{dx}{x} e^{-x/\xic} \cdot \lprox = \frac{2\pi^2 \nu_{\rm M}(\partial_y \Phi)^4 t^4 p_F^2}{v^2 D}\ln{\frac{\xic}{l_{\rm mfp}}}\cdot \lprox. 
\end{equation}
In the latter equality, we regularized the logarithmic divergence at small distances by the mean free path $l_{\rm mfp}$, i.e., by the length scale at which the diffusive behavior ceases. 

Finally, it is convenient to express the factor in front of the logarithm in Eq.~\eqref{eq:finalish} in terms of the normal-state conductivity of the metal $\sigma = 2e^2\nu_{\rm M}D$, and of the conductance per unit length of the interface $g = 2\pi^2 G_Q t^2 (\partial_y \Phi)^2 \nu_{\rm QH} \nu_{\rm M} p_F$ (here $\nu_{\rm QH} = (2\pi v)^{-1}$ is the density of edge states per spin projection and $G_Q = e^2 / \pi$). In this way we obtain Eq.~(8) of the main text.

\section{Derivation of $\langle \Theta^2 \rangle$ for the forward scattering phase $\Theta$}

Here we present the derivation of $\langle\Theta^2\rangle$ for the forward scattering phase $\Theta$ accumulated by an electron across a short proximitized segment.
We can obtain an expression for $\Theta$ similarly to how we found the amplitude $\he$, see Eq.~(7) of the main text. By treating $H_{\rm prox}$ in Eq.~(3) as a perturbation, we find:
\begin{equation}
    \Theta = \frac{(\partial_y \Phi)^2 t^2}{v}\hspace{-0.1cm}\int_0^\lprox \hspace{-0.1cm} dx_1 dx_2 e^{i\kf (x_1 - x_2)}\partial^2_{y_1 y_2}{\cal G}_{\rm ee} (x_1, x_2).\label{eq:amplitudes2}
\end{equation}
Here ${\cal G}_{\rm ee}(x_1, x_2) = {\cal G}_{\rm ee}(\bm r_1, \bm r_2 | E = 0)|_{y_{1,2}, z_{1,2} = 0}$ is the normal component of the superconductor Green's function. 
It is easy to verify using Eq.~\eqref{eq:greeen} at $E = 0$ that $\langle \Theta \rangle = 0$.
An expression for $\langle \Theta^2 \rangle$ can be obtained similarly to how we found $\langle |\he|^2\rangle$. A counterpart of Eq.~\eqref{eq:var3} is 
\begin{align}
     \langle \Theta^2 \rangle = &\frac{2\pi \nu_{\rm M}(\partial_y \Phi)^4 t^4 p_F^2}{v^2}\int^{\lprox}_{0} dx_{1}dx_{2}dx_{3}dx_{4}e^{ik_\mu(x_1 -x_2)}e^{-ik_\mu(x_3 - x_4)}\int \frac{\epsilon d\epsilon}{\Delta^2 +\epsilon^2} \frac{\epsilon^\prime d\epsilon^\prime}{\Delta^2+\epsilon^{\prime 2}} \notag\\
    &\times{\rm Re}\,{\difz}(x_1,x_2|\epsilon - \epsilon^\prime)\,\big[\delta(x_1 -x_3) \delta(x_2-x_4) + \delta(x_1 -x_4)\delta(x_2-x_3)\big],
\end{align}
where we also used Eq.~\eqref{eq:vertex_expr} for functions $V$. 
Using Eq.~\eqref{eq:diffuson_expl} for ${\difz}(x_1,x_2|\epsilon - \epsilon^\prime)$, we can rewrite the above expression as
\begin{align}
    \langle \Theta^2 \rangle &= \frac{\pi^2 \nu_{\rm M}(\partial_y \Phi)^4 t^4 p_F^2}{2v^2 D} \int^{\lprox}_{0} \frac{dx_{1}dx_{2}}{|x_1 - x_2|}e^{-|x_1-x_2|/\xic} \big[1 + \cos[2k_{\mu} (x_1 - x_2)]\big].\label{eq:fwdfinal}
\end{align}
The distance between points $x_1$ and $x_2$ here does not exceed the coherence length $\xic$. The latter satisfies $\xic \ll l_B \lesssim |k_\mu|^{-1}$ for a type II superconductor in field $B \ll H_{\rm c2}$. These estimates mean that the argument of cosine in Eq.~\eqref{eq:fwdfinal} is small, allowing one to approximate $\cos[2k_\mu (x_1 - x_2)] = 1$. Then, the right hand side of Eq.~\eqref{eq:fwdfinal} becomes identical to that of Eq.~\eqref{eq:finalish} for $\langle |\he|^2\rangle$. As a result, we obtain
\begin{equation}\label{eq:Theta}
    \langle \Theta^2 \rangle = \frac{\lprox}{\lA},
\end{equation}
where $\lA$ is given by Eq.~(8) of the main text. 

\section{Derivation of the conductance correlation function\label{sec:condcorrf}}

Here we present the derivation of the conductance correlation function ${\cal C}(\delta n, \delta B) = \lngl G(n, B) \cdot G(n + \delta n, B + \delta B) \rngl$,
which we use to obtain Eqs.~(16) and (17) of the main text.

To start with, we briefly discuss the main mechanism leading to the loss of correlation between the values of $G$ at parameters $(n, B)$ and $(n + \delta n, B + \delta B)$, respectively. Firstly, the variation $(\delta n, \delta B$) shifts the Fermi momentum of chiral electrons by $\delta k_\mu(\delta n, \delta B)$. As discussed after Eq.~(15) of the main text, this affects the phases of the Andreev amplitudes $\alpha(x)$. The phases are also affected by the change in the diamagnetic current flowing along the superconductor's surface.
The two effects can be accounted for by adding the phase factor to the Andreev amplitude, $\alpha (x) \rightarrow \alpha(x) e^{2i\delta k^{\rm (tot)}_{\mu} x}$, where $\delta k^{\rm (tot)}_{\mu} = \delta k_\mu  - \frac{1}{2} \delta (\partial_x \varphi)$ and $\partial_x \varphi$ is the gradient of the order parameter phase associated with the diamagnetic current. The variation $(\delta n, \delta B)$ also affects the magnitudes $|\alpha(x)|$ and $|\vartheta(x)|$. 
The reason is the dependence of $\partial_y \Phi$ and $v$ in Eqs.~\eqref{eq:amplitudes} and \eqref{eq:amplitudes2} on $n$ and $B$.
The magnitudes change as $|\alpha(x)| \rightarrow (1 + \delta g / g) |\alpha(x)|$ and $|\vartheta(x)| \rightarrow (1 + \delta g / g) |\vartheta(x)|$, where we used the relation for $g$ presented at the end of Sec.~\ref{sec:lA}.

To find ${\cal C}(\delta n, \delta B)$, we use Eq.~(10) of the main text to compare the results of the wave function evolution across the proximitized segment at parameters $(n, B)$ and $(n+\delta n, B + \delta B)$. We denote the components of the wave function by $a_{\rm e}(x)$, $a_{\rm h}(x)$ and $b_{\rm e}(x)$, $b_{\rm h}(x)$ for the respective sets of parameters.
The corresponding evolution equations read
\begin{align}\label{eq:evol}
    i \frac{\partial}{\partial x}
    \begin{pmatrix}
    a_{\rm e}(x)\\
    a_{\rm h}(x)
    \end{pmatrix}&=
    \begin{pmatrix}
    -\vartheta(x) & \alpha^\star(x)\\
    \alpha(x) & \vartheta(x)
    \end{pmatrix}
    \begin{pmatrix}
    a_{\rm e}(x)\\
    a_{\rm h}(x)
    \end{pmatrix},\\
    i \frac{\partial}{\partial x}
    \begin{pmatrix}
    b_{\rm e}(x)\\
    b_{\rm h}(x)
    \end{pmatrix}&=
    \left(1 + \frac{\delta g}{g}\right)
    \begin{pmatrix}
    -\vartheta(x) & \alpha^\star(x) e^{-2i\delta k^{\rm (tot)}_{\mu} x}\\
    \alpha(x) e^{2i\delta k^{\rm (tot)}_{\mu} x} & \vartheta(x)
    \end{pmatrix}
    \begin{pmatrix}
    b_{\rm e}(x)\\
    b_{\rm h}(x)
    \end{pmatrix}.\label{eq:evol2}
\end{align}
We can represent ${\cal C}(\delta n, \delta B)$ in terms of the wave function components as
\begin{equation}\label{eq:Cintermsofwf}
{\cal C}(\delta n, \delta B) = \lngl G^2 \rngl \frac{\lngl |a_{\rm h}(\lprox)|^2 \cdot |b_{\rm h}(\lprox)|^2\rngl}{\lngl |a_{\rm h}(\lprox)|^2 \cdot |a_{\rm h}(\lprox)|^2\rngl}.
\end{equation}
To determine $\lngl |a_{\rm h}(\lprox)|^2 \cdot |b_{\rm h}(\lprox)|^2\rngl$, we derive
a system of differential equations describing the evolution with $x$ of the correlators  $\lngl a_i^\star(x) a_j(x) \cdot b_k^\star(x) b_l(x) \rngl$, where $i,j,k,l = {\rm e, h}$. In fact, a closed system of equations can be obtained using Eq.~(9) of the main text and following the approach described in Ref.~\cite{ovchinnikov1980}.
The system has a particularly simple form in terms of the following variables:
\begin{align}
c_0(x) &= \lngl |a_{\rm h}(x)|^2 \cdot |b_{\rm h}(x)|^2 \rngl + e^{-2\big(1+(1+\frac{\delta g}{g})^2\big) \frac{x}{\lA}} / 4,\label{eq:c0}\\
c_+(x) &= {\rm Re}\,\lngl a_{\rm e}^\star(x)a_{\rm h}(x) \cdot b_{\rm h}^\star (x)b_{\rm e}(x)\rngl,\label{eq:cp}\\
c_-(x) &= {\rm Im}\,\lngl a_{\rm e}^\star(x)a_{\rm h}(x) \cdot b_{\rm h}^\star (x)b_{\rm e}(x)\rngl.\label{eq:cm}
\end{align}
We obtain
\begin{equation}\label{eq:system}
\frac{\partial}{\partial x}\left(\begin{array}{c}
c_0(x)\\
c_+(x)\\
c_-(x)
\end{array}\right)=
\frac{1}{\lA}\left(\begin{array}{ccc}
-2\big( 1+\big(1+\frac{\delta g}{g}\big)^{2}\big)  & 2\big(1+\frac{\delta g}{g}\big) & 0\\
4\big(1+\frac{\delta g}{g}\big) & -\big( 1+\big(1 + \frac{\delta g}{g}\big)^{2}\big) -2\big(\frac{\delta g}{g}\big)^{2} & 2\delta k_\mu^{\rm (tot)}\lA\\
0 & -2\delta k_\mu^{\rm (tot)}\lA & -\big( 1+\big(1 + \frac{\delta g}{g}\big)^{2}\big) -2\big(\frac{\delta g}{g}\big)^{2}
\end{array}\right)\left(\begin{array}{c}
c_0(x)\\
c_+(x)\\
c_-(x)
\end{array}\right)
\end{equation}
(we also made a gauge transformation $b_{\rm e/h}(x) \rightarrow e^{\mp i\delta k_{\mu}^{\rm (tot)} x} b_{\rm e/h}(x)$ when deriving the system). The initial conditions are $c_0(0) = 1/4$, $c_\pm (0) = 0$. 

Let us assume that $\delta k_\mu^{(\rm tot)}\lA\ll 1$ and $\delta g / g \ll 1$. Under these conditions, system \eqref{eq:system} can be analyzed with the help of the perturbation theory. At $\delta g = 0$ and $\delta k_\mu^{(\rm tot)} = 0$, the $3\times 3$ matrix on the right hand side of  Eq.~\eqref{eq:system} has an eigenvalue $\omega = 0$. The zero eigenvalue corresponds to the steady state solution of the Fokker-Planck equation, see Eq.~(11) of the main text. The respective eigenvector is $(1, 2, 0)^T$.
The correction to $\omega = 0$ due to finite $\delta k_\mu^{\rm (tot)}$ and $\delta g$ is of the second order in these parameters:
\begin{equation}
    \omega = -\frac{4}{3} (\delta k_{\mu}^{(\rm tot)})^2 \lA - \frac{8}{3} \left(\frac{\delta g}{g}\right)^2 \frac{1}{\lA}.
\end{equation}
Using this expression, we find the solution of system \eqref{eq:system} at $x \gg \lA$:
\begin{equation}\label{eq:solforcs}
\left(\begin{array}{c}
c_0(x)\\
c_+(x)\\
c_-(x)
\end{array}\right)
=
\frac{1}{12} 
\begin{pmatrix}
1\\
2\\
0
\end{pmatrix}
\exp\Big[-\frac{4}{3} (\delta k_{\mu}^{(\rm tot)})^2 \lA x - \frac{8}{3} \Big(\frac{\delta g}{g}\Big)^2 \frac{x}{\lA}\Big].
\end{equation}
Setting $x = \lprox$ and using Eqs.~\eqref{eq:c0} and \eqref{eq:Cintermsofwf}, we obtain
\begin{equation}\label{eq:corrgen}
    {\cal C}(\delta n, \delta B) = \lngl G^2 \rngl \exp \Big[-\frac{4}{3} (\delta k_{\mu}^{(\rm tot)})^2 \lA \lprox -  \frac{8}{3} \left(\frac{\delta g}{g}\right)^2 \frac{\lprox}{\lA}\Big].
\end{equation}

We now apply the general result \eqref{eq:corrgen} to find the conductance correlation function with density ${\cal C}(\delta n)$. The change of the wave vector $k_\mu$ upon the variation $\delta n$ can be expressed as (we recall that $\hbar = 1$)
\begin{equation}\label{eq:kmudn}
    \delta k_\mu = \frac{\partial \mu}{\partial n} \frac{\delta n}{v},
\end{equation}
where $\partial \mu / \partial n$ is the inverse compressibility of the quantum Hall state.
The influence of $\delta n$ on $g$ can be disregarded provided $\lA \gg l_B = \sqrt{c / eB}$. Assuming the latter condition to be satisfied, we disregard the second term in the square brackets of Eq.~\eqref{eq:corrgen}. Then, substituting Eq.~\eqref{eq:kmudn} in Eq.~\eqref{eq:corrgen} we arrive to Eqs.~(16) and (17) of the main text.

\section{Derivation of Eq.~(18) for ${\cal C}_{\rm jump}(d)$}

In this section, we present details of the derivation of Eq.~(18) for the variance of the conductance jumps ${\cal C}_{\rm jump}(d) = \langle (\delta G)^2 \rangle$. We assume the proximitized segment to be long throughout the section, $\lprox \gg \lA$.

To find ${\cal C}_{\rm jump}(d)$, we compare the results of the wave function evolution across the proximitized segment before and after a vortex has entered the superconductor.
We denote the wave function components as $a_{\rm e}(x)$ and $a_{\rm h}(x)$ before the vortex entrance, and as $b_{\rm e}(x)$ and $b_{\rm h}(x)$ after it. The corresponding evolution equations are given by
\begin{equation}\label{eq:evol_v}
    i \frac{\partial}{\partial x}
    \begin{pmatrix}
    a_{\rm e}(x)\\
    a_{\rm h}(x)
    \end{pmatrix}
    = 
    \begin{pmatrix}
    -\vartheta(x) & \alpha^\star(x)\\
    \alpha(x) & \vartheta(x)
    \end{pmatrix}
    \begin{pmatrix}
    a_{\rm e}(x)\\
    a_{\rm h}(x)
    \end{pmatrix}, \quad \quad
    i \frac{\partial}{\partial x}\begin{pmatrix}
    b_{\rm e}(x)\\
    b_{\rm h}(x)
    \end{pmatrix} = 
    \begin{pmatrix}
    -\vartheta(x) & \alpha^\star(x) e^{i\delta \varphi(x)}\\
    \alpha(x)e^{-i\delta\varphi(x)} & \vartheta(x)
    \end{pmatrix} 
    \begin{pmatrix}
    b_{\rm e}(x)\\
    b_{\rm h}(x)
    \end{pmatrix}.
\end{equation}
Here $\delta \varphi(x) = \pi + \arctan ([x - x_{\rm v}] / d)$ is the phase induced by the entered vortex (we assume that pinning in the superconductor is strong enough so that the entrance of the vortex does not affect the preexisting vortex distribution). As mentioned in the main text, $d$ is the distance between the vortex core and the interface, and $x_{\rm v}$ is the core's coordinate along the $x$-direction.

The variance of the conductance jumps can be expressed in terms of the wave functions components as:
\begin{equation}\label{eq:jump_def}
    {\cal C}_{\rm jump}(d) = 2 \lngl G^2 \rngl \left[1 - \frac{\lngl |a_{\rm h}(\lprox)|^2 \cdot |b_{\rm h}(\lprox)|^2\rngl}{\lngl |a_{\rm h}(\lprox)|^2 \cdot |a_{\rm h}(\lprox)|^2\rngl}
\right].
\end{equation}
To find $\lngl |a_{\rm h}(\lprox)|^2 \cdot |b_{\rm h}(\lprox)|^2\rngl$, we derive a system of equations for correlators $\lngl a_i^\star(x) a_j(x) \cdot b_k^\star(x) b_l(x) \rngl$ similarly to how we did it in Sec.~\ref{sec:condcorrf}. The system reads
\begin{equation}\label{eq:sys_2}
    \frac{\partial}{\partial x}\left(\begin{array}{c}
c_0(x)\\
c_+(x)\\
c_-(x)
\end{array}\right)
=
\frac{1}{l_{\rm A}}\left(\begin{array}{ccc}
-4  & 2 & 0\\
4 & -2  & -\lA \partial_x \delta \varphi(x) \\
0 & \lA \partial_x \delta \varphi(x) & -2 
\end{array}\right)\left(\begin{array}{c}
c_0(x)\\
c_+(x)\\
c_-(x)
\end{array}\right).
\end{equation}
Here variable $c_{0}(x) = \lngl |a_{\rm h}(x)|^2 \cdot |b_{\rm h}(x)|^2 \rngl + e^{-4 x / \lA} / 4$, whereas variables $c_\pm (x)$ are defined in the same way as in Eqs.~\eqref{eq:cp} and \eqref{eq:cm}. 

System of equations \eqref{eq:sys_2} can be solved analytically in the two limiting cases, $d \ll \lA$ and $d \gg \lA$. Let us start with the former case. The condition $d \ll \lA$ means that the kink in the superconducting phase $\delta \varphi(x)$ is narrow. This suggests one to approximate
\begin{equation}\label{eq:deltaappr}
    \partial_x \delta \varphi(x) = 2\pi\cdot  \frac{1}{\pi}\frac{d}{d^2 + (x - x_{\rm v})^2}\approx 2\pi \delta (x - x_{\rm v})
\end{equation}
in Eq.~\eqref{eq:sys_2}.
However, such an approximation is too crude. Indeed, it can be easily verified that the vector $(c_0(x),c_+(x),c_-(x))^T$ does not change across $x = x_{\rm v}$ if we replace $\partial_x \delta \varphi(x) \rightarrow 2\pi \delta (x - x_{\rm v})$. 
Thus, ${\cal C}_{\rm jump}(d)=0$ to the zeroth order in $d / \lA$.

The leading in $d / \lA$ result for ${\cal C}_{\rm jump}(d)$ can be obtained in the following way. First of all, we go to a rotating frame in Eq.~\eqref{eq:sys_2}:
\begin{equation}
    \left(\begin{array}{c}
c_0(x)\\
c_+(x)\\
c_-(x)
\end{array}\right) = \exp\Big[ \begin{pmatrix}
0 & 0 & 0\\
0 & 0 & -\delta\varphi(x)\\
0 & \delta \varphi(x) & 0
\end{pmatrix}\Big]\left(\begin{array}{c}
\tilde{c}_0(x)\\
\tilde{c}_+(x)\\
\tilde{c}_-(x)
\end{array}\right)
\end{equation}
(we choose the frame in such a way that terms $\propto \partial_{x}\delta \varphi(x)$ cancel on two sides of the equation after the transformation).
In this frame, the ``scatterer'' associated with the vortex is described by a local perturbation of magnitude $\sim 1 / \lA$ and width $\sim d$. It can be treated using an analog of Born approximation. A straightforward calculation leads to
\begin{equation}
\lngl |a_{\rm h}(\lprox)|^2 \cdot |b_{\rm h}(\lprox)|^2\rngl = \lngl |a_{\rm h}(\lprox)|^2 \cdot |a_{\rm h}(\lprox)|^2\rngl \left(1 - \frac{16\pi}{3}\frac{d}{\lA}\right).
\end{equation}
Using this expression in Eq.~\eqref{eq:jump_def}, we obtain the result presented in the first line of Eq.~(18) of the main text.

Now we consider the limit of $d \gg \lA$. In this limit, we can account for $\partial_x \delta \varphi(x) \lA$ in system \eqref{eq:sys_2} with the help of the adiabatic approximation. Using the similarity of system \eqref{eq:sys_2} to system \eqref{eq:system} (taken at $\delta g = 0$), we find by generalizing Eq.~\eqref{eq:solforcs}:
\begin{equation}
\left(\begin{array}{c}
c_0(x)\\
c_+(x)\\
c_-(x)
\end{array}\right) \approx \frac{1}{12} \begin{pmatrix}
1 \\ 2\\ 0
\end{pmatrix} e^{-\frac{1}{3}\int_0^x [\partial_x \delta \varphi(x^\prime) \lA]^2 dx^\prime}.
\end{equation}
Taking $x = \lprox \gg \lA$, computing the integral in the exponent, and using the definition of a variable $c_0(x)$, we find
\begin{equation}
    \lngl |a_{\rm h}(\lprox)|^2 \cdot |b_{\rm h}(\lprox)|^2\rngl   = \lngl |a_{\rm h}(\lprox)|^2 \cdot |a_{\rm h}(\lprox)|^2\rngl\,\exp\Big[-\frac{2\pi \lA}{d}\Big] \approx \lngl |a_{\rm h}(\lprox)|^2 \cdot |a_{\rm h}(\lprox)|^2\rngl  \left(1 -\frac{2\pi \lA}{d}\right).
\end{equation}
Substituting this expression in Eq.~\eqref{eq:jump_def}, we obtain the result presented in the second line of Eq.~(18) of the main text.